\documentclass[aps,prd,preprint,groupedaddress,showpacs]{revtex4-2}
\usepackage{amsmath}
\usepackage{bm}
\usepackage{braket}
\usepackage{color}
\usepackage{graphicx}
\usepackage{hyperref}
\usepackage{amssymb}
\usepackage{dcolumn}
\usepackage{multirow}

\begin{document}

\title{Effect of isoscalar and isovector scalar fields on baryon semileptonic decays in nuclear matter}
\author{Koichi Saito}
\email[]{koichi.saito@rs.tus.ac.jp}
\affiliation{Department of Physics and Astronomy, Tokyo University of Science, Noda 278-8510, Japan}
\author{Tsuyoshi Miyatsu}
\email[]{tsuyoshi.miyatsu@ssu.ac.kr}
\affiliation{Department of Physics and OMEG Institute, Soongsil University, Seoul 06978, Republic of Korea}
\author{Myung-Ki Cheoun}
\email[]{cheoun@ssu.ac.kr}
\affiliation{Department of Physics and OMEG Institute, Soongsil University, Seoul 06978, Republic of Korea}
\date{\today}

\begin{abstract}
The precise determination of the Cabibbo-Kobayashi-Maskawa (CKM) matrix elements is very important, because it could be a clue to new physics beyond Standard Model. This is particular true of $V_{ud}$, because it is the main contribution to the unitary condition of the CKM matrix elements. The level of accuracy for the test of the unitarity involving the element $V_{ud}$ is now of the order of $10^{-4}$.  Because the precise data for $V_{ud}$ is usually extracted from super-allowed nuclear $\beta$ decay, it is quite significant to investigate the breaking of SU(3) flavor symmetry on the weak vector coupling constant in nuclear matter. The purpose of this paper is to investigate how the isoscalar scalar ($\sigma$) and the isovector scalar ($\delta$ or $a_0$) mean-fields affect the weak vector and axial-vector coupling constants for semileptonic baryon (neutron, $\Lambda$ or $\Xi^-$) decay in asymmetric nuclear matter.  To do so, we use the quark-meson coupling (QMC) model, where nuclear matter consists of nucleons including quark degrees of freedom bound by the self-consistent exchange of scalar and vector mesons. We pay careful attention to the center of mass correction to the quark currents in matter.  We then find that, for neutron $\beta$ decay in asymmetric nuclear matter, the defect of the vector coupling constant due to the $\delta$ field can be of the order of $10^{-4}$ at the nuclear saturation density, which is the same amount as the level of the current uncertainty in the measurements.  It is also interesting that, in neutron-rich matter, there exists a certain low density at which isospin symmetry is restored, that is, the $u$-$d$ quark mass difference vanishes.  Furthermore, we qualitatively investigate the variation of axial-vector coupling constants in matter, to which the $\sigma$ field mainly contributes. We conclude that the effect of the isoscalar scalar and the isovector scalar fields should be considered in baryon semileptonic decays in nuclei. 
\end{abstract}

\maketitle

\newpage

\section{Introduction} \label{sec:introduction}

It is very important to understand baryon semileptonic decays precisely, because it could be a probe for new physics beyond Standard Model~\cite{brodeur23nuclear, hayen24opp}.  In particular, testing the unitarity of the Cabibbo-Kobayashi-Maskawa (CKM) matrix is one of the more important issues of this problem.  The main information on the largest, top-left corner element of the CKM matrix is now obtained from super-allowed nuclear $\beta$ decays with an impressive precision as $V_{ud} = 0.97373 \pm 0.00031$~\cite{pdg}.  To extract such precise value from the experimental measurements, it is necessary to estimate the effects of nuclear isospin-breaking as well as a number of small nuclear structure corrections up to the level of $10^{-4}$.  

Until now the nuclear corrections in the determination of the quark-level vector coupling, $V_{ud}$, have been calculated by the conventional nuclear theory with point-like nucleons~\cite{hardy20super, miller23str, gorchtein24super}.  Of course, for the nucleon itself there has been considerable investigation of the effect on the vector form factor of the breaking of conserved vector current (CVC) generated by the quark mass difference $\Delta_{ij} = m_i - m_j$ ($i$ or $j$ denotes quark flavor).  The Behrends-Sirlin-Ademollo-Gatto (BSAG) theorem~\cite{behrends60eff,ademollo64nonr} tells us that any correction to CVC must be of second order in $\Delta_{ij}$.  Then, Behrends and Sirlin~\cite{behrends60eff}  concluded that the correction to the effective vector $\beta$ decay matrix element due to the $u$-$d$ mass splitting is typically of the order $(\Delta_{np}/M_N)^2 \simeq (\Delta_{du}/M_N)^2 \simeq (1.3/940)^2 \simeq 2 \times 10^{-6}$, where $\Delta_{np}$ is the mass difference between proton ($M_p$) and neutron ($M_n$), and $M_N$ is the average of those masses.  This amount is quite small and thus seems negligible.  A similar effect occurs in the case of the axial-vector current, to which, however, the BSAG theorem is not applied, and hence the breaking starts from first order in $\Delta_{ij}$. 
In Refs.~\cite{donoghue82quark, eich85static, barik85weak, donoghue87km, schlumpf95beta, mendieta98su3, cabibbo04semi, yamanishi07fd, ledwig08semi, sharma09weak, yang15hy, zhang24hy}, the effects of flavor SU(3) breaking on semileptonic baryon decays and baryon properties have been studied using various quark models. 

Recently, Crawford and Miller~\cite{crawford22charge} have discussed in detail the effect of charge-symmetry-breaking (CSB) on neutron $\beta$ decay using the nonrelativistic quark model with a harmonic oscillator potential.  They have reported that CSB, which comes from the corrections of kinetic energy, electromagnetic interaction and gluon-exchange interaction,  decreases the value of the vector matrix element of $\beta$ decay by a factor of about $10^{-4}$.  This amount is about $100$ times larger than the previous estimate by Behrends and Sirlin, and is of the order of the current uncertainty in the measurements.  This indicates that the effect of baryon substructure should be included in the analyses of semileptonic baryon decays. 

Furthermore, it is very important to examine at the quark level how nuclear binding affects the $\beta$ decay of baryon itself {\em in nuclei}.   In order to carry out such study, one needs a model of nuclear structure involving explicit quark degrees of freedom which can provide an acceptable description of various nuclear-matter properties.  The quark-meson coupling (QMC) model~\cite{guichon88a, saito94a, saito07nucleon, guichon18qmc, krein18nucl}, in which nuclear matter consists of non-overlapping nucleon bags bound by the self-consistent exchange of $\sigma$ and $\omega$ mesons, seems very suited to the problem.  

In Ref.~\cite{saito95fermi}, in fact, the QMC model was applied to investigate the effect of isospin breaking on the Fermi decay constant of the nucleon in nuclear matter.  Then, it was shown that the difference between the bag radii of proton and neutron, which comes from $\Delta_{du}$, leads to a large correction to the Fermi coupling that is {\em inconsistent} with the BSAG theorem.  In the transition matrix elements involving {\em inelastic} processes like the weak baryon decays, there is some arbitrariness caused by the disparity of bag radii.  This trouble was already pointed out in earlier works~\cite{donoghue75phen}. 
To avoid this radius problem in the bag model, perturbation theory is very powerful and the expansion can be performed in terms of $\Delta_{du}/(\omega_\alpha-\omega_0)$, where $\omega_{0(\alpha)}$ is the ground (radial excited) state energy of a quark in the bag.  Then, Guichon et al.~\cite{guichon11fermi} have reexamined the renormalization for the vector current in nuclear matter, and have found that the quark mass difference and the pion mass difference generate a correction of order $10^{-4}$, which is close to the result in Ref.~\cite{crawford22charge} and is again much larger than the estimation by Behrends and Sirlin~\cite{behrends60eff}.  

In the light of a recent proposal at J-PARC to measure the vector and axial charges in the strangeness changing $\beta$ decay of a bound $\Lambda$ in hypernuclei~\cite{kamada22exp}, the variation of vector and axial-vector couplings for $\Lambda$ beta-decay in matter has been discussed in the QMC model~\cite{guichon17Lamb}.  Here, one of the advantages of hypernuclei is that one can identify the location of decaying hyperon in a nucleus, namely either near the center of the nucleus or outside with lower density. This is a  fascinating feature. However, it is necessary to be careful in such cases because SU(3) breaking due to the $u$-$s$ quark mass difference $\Delta_{su}$ could be much larger than in the isospin breaking case.  Therefore, a more detailed estimate of the in-medium SU(3) breaking at the quark level is very interesting and anticipated. 

The present paper is aimed at investigating the variation of weak vector and axial-vector coupling constants in asymmetric nuclear matter.  In matter, it is expected that the isoscalar scalar ($\sigma$) and isovector scalar ($\delta$ or $a_0$) mesons change the $u$ and $d$ quark masses from the values in vacuum~\cite{saito94ns, guichon11fermi}, in which the effect of isovector scalar field is less discussed so far, and that the weak couplings in matter are also modified by such variation.  In this paper we thus focus our discussion on two issues: one is the role of $\sigma$ and $\delta$ mesons in semileptonic decays of baryons (for example, neutron, $\Lambda$ and $\Xi^-$), the other is the center of mass (cm) correction to the quark current in nuclear matter.  In order to investigate those effects on the weak couplings in detail, we do not explicitly evaluate the other contributions to SU(3) breaking like the effects of pionic cloud and gluon-exchange interactions, but we treat them as parameters.  The full calculation including such contributions is a future consideration. 

In the QMC model, baryons in matter are usually described by the MIT bag model.  However, in the present calculation, instead of the bag, we would like to adopt an alternative way to confine the quarks, that is, a relativistic confinement potential of the scalar-vector harmonic oscillator (HO) type 
\begin{equation}
  U(r) = \frac{c}{2} (1+\gamma_0) r^2 . 
  \label{eq:conpot}
\end{equation}
The QMC model with the relativistic HO potential is sometimes referred as the quark mean field (QMF) model~\cite{toki98qmf, shen00qmf} or the modified quark-meson coupling (MQMC) model~\cite{barik13n}.  This quark model has some pleasant feature that all quantities of interest can be basically calculated analytically, and that the model itself does not involve the radius problem, which occurs in the bag model, and hence is still applicable to the case where SU(3) breaking is not small.  Of course, it has also some disadvantages.  It is well known that the scalar-vector type potential does not produce any spin-orbit splitting in the baryon spectra. In particular, concerning the axial-vector coupling constant, because such a quark model requires that the quark has a constituent quark mass ($\sim 300$ MeV), it may be difficult to reproduce the experimental value itself~\cite{chodos74bag, cbm81, tonyadv13}.  However, we are interested primarily in SU(3) breaking, not in the absolute value of axial charge predicted within a given quark model.  Thus, it is possible to discuss how the $\sigma$ and $\delta$ fields contribute to the in-medium axial-vector coupling {\em qualitatively}. 

Of course, a more accurate approach from first principles, such as lattice QCD, would be required to analyse the precise experimental data in the future.  At present, in Refs.~\cite{sasaki09lat, erkol10ax, sasaki12lat, sasaki17lat, savanur20lat, seng23quark} the weak form factors of nucleon and hyperons have been discussed in terms of lattice QCD. 

Here is an outline. Section~\ref{sec:quark-model} introduces the relativistic quark model to describe the properties of baryons.  Then, we calculate the weak vector and axial-vector coupling constants in vacuum in Sec.~\ref{sec:weakff}.  We introduce the QMC model and discuss the properties of asymmetric nuclear matter in Sec.~\ref{sec:qmcmodel}.  The weak coupling constants in matter are then estimated and discussed in Sec.~\ref{sec:weakmatter}. 
Finally, we give summary and discussion in Sec.~\ref{sec:sumconc}.  In Appendices, several useful formulae are summarized. 

\newpage

\section{Relativistic quark model} \label{sec:quark-model}

In the present paper, we adopt a relativistic confinement potential of the scalar-vector harmonic oscillator (HO) type to confine quarks, which is given by Eq.~(\ref{eq:conpot}).  We assume that the strength parameter, $c$, respects flavor symmetry.  
It is well known that the Dirac equation with the scalar-vector type HO potential can be solved analytically~\cite{ferreira771, ferreira772}.  In free space, the lowest-state solution for a quark with flavor $i \, (= u, d, s)$ is given by
\begin{align}
\psi_i({\vec r}\, ) 
  & = \frac{N_i}{\sqrt{4\pi} a_i} 
  \begin{pmatrix}
1 \\
i {\vec \sigma} \cdot {\hat r} \frac{1}{\lambda_i a_i} \left( \frac{r}{a_i} \right) 
\end{pmatrix}
  e^{-r^2/2a_i^2} \chi_i ,  \label{eq:sol} \\
  N_i^2 & = \frac{8}{\sqrt{\pi} a_i} \frac{\lambda_i^2 a_i^2}{2\lambda_i^2 a_i^2+3},  \ \ \ 
  \lambda_i = \epsilon_i + m_i, \ \ \  a_i^2 = \frac{1}{\sqrt{c\lambda_i}} = \frac{3}{\epsilon_i^2-m_i^2} ,    \label{eq:norm}
\end{align}
where $m_i$ is the constituent quark mass, $a_i$ is the length scale and $\chi_i$ is the quark spinor.  Here, the single-particle quark energy, $\epsilon_i$, is determined by
\begin{equation}
  \sqrt{\epsilon_i + m_i} (\epsilon_i - m_i) = 3 \sqrt{c} .
  \label{eq:qenergy}
\end{equation}

The zeroth-order energy of the low-lying baryon, $E_B^0$, is then simply given by a sum of the quark energies, $E_B^0 = \sum_i \epsilon_i$.  
Now we should take into account some corrections to the baryon energy such as the spin correlations, $E^{spin}_B$, due to the quark-gluon and quark-pion interactions~\cite{barik86mass, zhu23eos}, and the cm correction, $E^{cm}_B$.  Here we do not calculate the spin correlation explicitly, but we assume that it can be treated as a constant parameter which is fixed so as to reproduce the baryon mass~\cite{toki98qmf, shen00qmf}.  The cm correction to the spurious motion can be calculated analytically (for detail, see Appendix~\ref{app:mass}).  In the present calculation, we shall consider four baryons, namely  proton, neutron, $\Lambda$ and $\Xi^-$.  Then, we find the cm correction to the energy as 
\begin{align}
  E^{cm}_{p}
  & =\frac{11}{3}(2A_{u}+A_{d})-\frac{2}{9}(C_{u}D_{u}+C_{u}D_{d}+C_{d}D_{u}),  \label{eq:cmcorrection1} \\
  E^{cm}_{n}
  & =\frac{11}{3}(A_{u}+2A_{d})-\frac{2}{9}(C_{u}D_{d}+C_{d}D_{u}+C_{d}D_{d}), \\
  E^{cm}_{\Lambda}
  & =\frac{11}{3}(A_{u}+A_{d}+A_{s})-\frac{1}{9}(C_{u}D_{d}+C_{u}D_{s}+C_{d}D_{u}+C_{d}D_{s}+C_{s}D_{u}+C_{s}D_{d}) ,  \\
  E^{cm}_{\Xi^-}
  & =\frac{11}{3}(A_{d}+2A_{s})-\frac{2}{9}(C_{s}D_{s}+C_{s}D_{d}+C_{d}D_{s}) , \label{eq:cmcorrection4}
\end{align}
where $A_i$, $C_i$ and $D_i$ are defined in Eq.~(\ref{eq:acd1}). 
The baryon mass is thus given by 
\begin{equation}
  M_B = E_B^0 + E^{spin}_B - E^{cm}_B .
  \label{eq:bmass}
\end{equation}

The mean-square charge radius of baryon is also given by Eq.~(\ref{eq:chargeradius}) in  Appendix~\ref{app:radius}.  We find that the charge radii including the cm correction are  
\begin{align}
  \langle r^2 \rangle_{p} &=D_{u} = \frac{11\epsilon_{u}+m_{u}}{\left(3\epsilon_{u}+m_{u}\right)\left(\epsilon_{u}^2-m_{u}^2\right)} ,   \ \ \ 
    \langle r^2 \rangle_{n} =\frac{1}{3} \left( D_{u} - D_d \right) , \label{eq:radiusp}  \\
    \langle r^2 \rangle_{\Lambda} &=\frac{1}{6} \left( 2D_{u} - D_d - D_s \right) , \ \ \ 
    \langle r^2 \rangle_{\Xi^-} =-\frac{1}{3} \left( D_{d} + 2D_s \right) . \label{eq:radiusb} 
\end{align}

In this model,  we have eight parameters: $c$, $m_u$, $m_d$, $m_s$, $E_p^{spin}$, $E_n^{spin}$, $E_\Lambda^{spin}$, $E_{\Xi^-}^{spin}$.  In order to reduce the number of parameters, we  assume $E_N^{spin}=E_p^{spin}=E_n^{spin}$, and consider the three cases: \\
case 1; $m_u = 250$ MeV, $m_s = 450$ MeV,  \\
case 2; $m_u = 300$ MeV, $m_s = 500$ MeV,  \\
case 3; $m_u = 350$ MeV, $m_s = 550$ MeV.  \\
The remaining five parameters, $c$, $m_d$, $E_N^{spin}$, $E_\Lambda^{spin}$ and $E_{\Xi^-}^{spin}$, are determined so as to reproduce the proton charge radius $\langle r^2 \rangle_p^{1/2} = 0.841$ fm~\cite{pdg} and the baryon masses: $M_p=937.64$ MeV, $M_n=939.70$ MeV, $M_\Lambda=1115.68$ MeV and $M_{\Xi^-} = 1321.71$ MeV.  Here, the electromagnetic self-energy for proton or neutron is subtracted from the observed  mass~\cite{saito94ns}.  

In Table~\ref{tab:parameters}, the parameters and the quark energies are summarized.  The properties of proton, neutron, $\Lambda$ and $\Xi^-$ hyperons in vacuum are also presented in Table~\ref{tab:baryons}. 

\newpage

\begin{table*}[h!]
  \caption{\label{tab:parameters}
    Parameters and quark energies in vacuum.
  }
  \begin{ruledtabular}
    \begin{tabular}{ccccccccccc}
   & $m_{u}$\footnote{Input} & $m_{d}$ & $m_{s}^{\text{ a}}$ & $\epsilon_{u}$ & $\epsilon_{d}$ & $\epsilon_{s}$ & $E_{N}^{spin}$ & $E_{\Lambda}^{spin}$ & $E_{\Xi^-}^{spin}$ & $c$  \\
     \cline{2-7} \cline{8-10} \cline{11-11}  
 case  &   \multicolumn{6}{c}{(MeV)} & \multicolumn{3}{c}{(MeV)} & (fm$^{-3}$)  \\
   \cline{1-1}     \cline{2-11}
 1 &  250 & 252.63 & 450 & 493.1 & 495.0 & 649.8 & $-236.8$ & $-227.7$ & $-190.2$ & 0.635 \\
 2 &  300 & 302.48 & 500 & 517.8 & 519.7 & 681.2 & $-334.5$ & $-331.3$ & $-299.8$ & 0.561 \\
 3 & 350 & 352.38 & 550 & 546.4 & 548.3 & 715.3 & $-441.9$ & $-443.3$ & $-416.5$ & 0.500 \\
    \end{tabular}
  \end{ruledtabular}
\end{table*}
\begin{table*}[h!]
  \caption{\label{tab:baryons}
    Properties of proton, neutron, $\Lambda$ and $\Xi^-$ in vacuum.
  }
  \begin{ruledtabular}
    \begin{tabular}{ccccccccccccc}
      & $E_{n}^{0}$ & $E_{p}^{0}$ & $E_{\Lambda}^{0}$ & $E_{\Xi^-}^{0}$ & $E_{n}^{cm}$ & $E_{p}^{cm}$ & $E_{\Lambda}^{cm}$ & $E_{\Xi^-}^{cm}$ & $\braket{r^{2}}_{n}$ & $\braket{r^{2}}_{p}\footnote{Input} $ & $\braket{r^{2}}_{\Lambda}$ & $\braket{r^{2}}_{\Xi^-}$ \\
      \cline{2-5} \cline{6-9} \cline{10-13}
     case & \multicolumn{4}{c}{(MeV)} & \multicolumn{4}{c}{(MeV)} & \multicolumn{4}{c}{(fm$^2$)} \\
 \cline{1-1}     \cline{2-13}
      1 & 1483.1 & 1481.2 & 1637.9 & 1794.6 & 306.5 & 306.7 & 294.5 & 282.7 & 0.001 & 0.707 & 0.025 & $-0.609$ \\
      2 & 1557.2 & 1555.2 & 1718.7 & 1882.1 & 283.0 & 283.1 & 271.7 & 260.6 & 0.001 & 0.707 & 0.023 & $-0.616$ \\
      3 & 1643.0 & 1641.1 & 1810.0 & 1978.9 & 261.3 & 261.5 & 251.0 & 240.7 & 0.001 & 0.707 & 0.021 & $-0.623$ \\
    \end{tabular}
  \end{ruledtabular}
\end{table*}
\clearpage

\section{Weak coupling constants in vacuum} \label{sec:weakff}

The transition matrix element for the decay of an initial baryon $B_1$ to a final baryon $B_2$ and lepton $l$ with its antineutrino ${\bar \nu}_l$, namely $B_1 \to B_2 + l + {\bar \nu}_l$, is proportional to the matrix element of the baryon weak current $J^\mu(x)$, which consists of the vector and axial-vector currents, as 
\begin{equation}
  \braket{B_2| J^\mu(x) |B_1} = C \left[ \braket{B_2| V^\mu(x) |B_1} + \braket{B_2| A^\mu(x) |B_1} \right] , 
  \label{eq:weakcurrent}
\end{equation}
with 
\begin{align}
\braket{B_2| V^\mu(x) |B_1} & =  {\bar u}_2(p_2) \left[ f_1(q^2) \gamma^\mu + \cdots \right] u_1(p_1) e^{iq\cdot x}  ,  \label{eq:weakv} \\ 
\braket{B_2| A^\mu(x) |B_1} & = {\bar u}_2(p_2) \left[ g_1(q^2) \gamma^\mu \gamma_5  + \cdots \right] u_1(p_1) e^{iq\cdot x}  , 
\label{eq:weaka}
\end{align}
where $C = \cos \theta_c \, (\sin \theta_c)$ is the Cabibbo angle for the strangeness transition $\Delta S = 0 \,  (1)$, and $p_{1 (2)}$ is the four-momentum of baryon $1 (2)$.  Here, $q^2=(p_1 - p_2)^2$ is the momentum transfer squared, which is generally quite small for any such decays. The ellipsis in Eq.~(\ref{eq:weakv}) or (\ref{eq:weaka}) refers to additional vector or axial-vector currents, which contribute to the decay rates only in order $q/M$ or higher.  Therefore, we here focus on the leading form factors $f_1(0)$ (vector coupling constant) and $g_1(0)$ (axial-vector coupling constant), and only consider the $\mu=0$ component of $V^\mu$ and the $\mu=3$ component of $A^\mu$ (all baryons are treated with spin up).  

The $\beta$ decay of the octet baryon, $B_1 \to B_2 + e^- + {\bar \nu}_e$, can be interpreted as the quark $\beta$ decay like $q_i \to q_j + e^- + {\bar \nu}_e$ inside the baryon, where $j=u$ and $i$ could be $d$ or $s$ quark.  The weak coupling constants without the cm correction, $f_1^{(0)}$ and $g_1^{(0)}$, are then calculated by the quark model  
\begin{equation}
   f_1^{(0)}(B_1 \to B_2) = \int d{\vec r} \braket{B_2| V^0(x) |B_1} ,  \ \ \ 
   g_1^{(0)}(B_1 \to B_2) = \int d{\vec r} \braket{B_2| A^3(x) |B_1} , 
  \label{eq:uncorcc}
\end{equation}
and they are rewritten by Eq.~(\ref{eq:sol})
\begin{align}
f_1^{(0)}(B_1 \to B_2) & = f_1^{\rm SU(3)} \times \int d{\vec r} \, {\bar \psi}_j({\vec r}) \gamma_0 \psi_i({\vec r})    \nonumber \\
& =  f_1^{\rm SU(3)} \times 2^{5/2} \left( \frac{a_i a_j}{a_i^2 + a_j^2} \right)^{3/2} \frac{\lambda_i a_i \lambda_j a_j + 3 a_i a_j /(a_i^2 + a_j^2)}{\sqrt{2 \lambda_i^2 a_i^2 + 3} \sqrt{2 \lambda_j^2 a_j^2 + 3}}   \nonumber \\
& \equiv  f_1^{\rm SU(3)} \times (1 - \delta f_1^{(0)}) ,  
\label{eq:gv0} 
\end{align}
and 
\begin{align}
g_1^{(0)}(B_1 \to B_2) & = g_1^{\rm SU(3)} \times \int d{\vec r} \, {\bar \psi}_j({\vec r}) \gamma^3 \gamma_5 \psi_i({\vec r})  ,  \nonumber \\
& = g_1^{\rm SU(3)} \times 2^{5/2} \left( \frac{a_i a_j}{a_i^2 + a_j^2} \right)^{3/2} \frac{\lambda_i a_i \lambda_j a_j - a_i a_j /(a_i^2 + a_j^2)}{\sqrt{2 \lambda_i^2 a_i^2 + 3} \sqrt{2 \lambda_j^2 a_j^2 + 3}} \nonumber \\
& \equiv g_1^{\rm SU(3)} \times (1 - \delta g_1^{(0)}) ,   
\label{eq:ga0}
\end{align}
where in both cases the superscript ${}^{\rm SU(3)}$ indicates the usual value obtained in exact {\rm SU(3)}~\cite{cabibbo03su3}.  
Now, using Eq.~(\ref{eq:qenergy}), 
we can easily verify that the vector coupling $f_1^{(0)}$ obeys the BSAG theorem, namely $\delta f_1^{(0)} = {\cal O}(\Delta_{ij}^2)$.  Here we should notice that the deviation from unity, $\delta f_1^{(0)} (\geq 0)$, is caused by the violation of SU(3) flavor symmetry, while $\delta g_1^{(0)}(\geq 0)$ is mainly generated by the relativistic effect of quark wave function, that is, the lower component of the wave function in Eq.~(\ref{eq:sol}).  

The numerical results for some semileptonic baryon decays in vacuum are shown in Table~\ref{tab:weakcc-vacuum}.  
For example, in case 2, we find that $\delta f_1^{(0)} \simeq 2.3 \times 10^{-6}$ and $\delta g_1^{(0)} \simeq 0.16$ in the neutron $\beta$ decay.  It is not surprising that $\delta f_1^{(0)}$ is of the order of $10^{-6}$, because the present calculation does not involve SU(3) breaking due to the effect of gluons and pions~\cite{crawford22charge}.  In contrast, we find that $\delta f_1^{(0)} \simeq 0.01$ and $\delta g_1^{(0)} \simeq 0.13$ in the $\Lambda$ or $\Xi^-$ decay.  Here, because of $\Delta_{su} \gg \Delta_{du}$, $\delta f_1^{(0)}$ in the hyperon decay is much larger than that in the neutron decay.  On the other hand, $\delta g_1^{(0)}$ in the hyperon decay is smaller than that in the neutron decay, because the $d$ quark is lighter ({\em more relativistic}) than the $s$ quark and thus the lower component of the $s$-quark wave function in hyperon is smaller than that of the $d$-quark wave function in neutron.  

Next we must add the cm correction to the uncorrected, coupling constants, $f_1^{(0)}$ and $g_1^{(0)}$.  To do so, the better way may be to use the method of wave packets proposed by Donoghue and Johnson~\cite{donoghue80pion}, because the matrix elements of weak currents are built out of quarks only~\cite{detar81qbm, bartelski84cm}.  For detail, see Appendix~\ref{app:current}. 
From Eqs.~(\ref{eq:corgv}) and (\ref{eq:corga}), we have 
\begin{equation}
 f_1 = \frac{f_1^{(0)}}{\rho_V} \equiv f_1^{\rm SU(3)} \times (1 - \delta f_1) ,  \ \ \ 
 g_1 = \frac{g_1^{(0)}}{\rho_A} \equiv g_1^{\rm SU(3)} \times (1 - \delta g_1)   ,
 \label{eq:corgva}  
\end{equation}
where $f_1$ and $g_1$ are respectively the corrected, vector and axial-vector coupling constants, and the cm corrections, $\rho_V$ and $\rho_A$, are given by Eqs.~(\ref{eq:exrhov}) and (\ref{eq:exrhoa}).  

We also present the numerical results in Table~\ref{tab:weakcc-vacuum}.  Because $\rho_V$ is, roughly speaking, estimated by the superposition of the normalization of baryon transition amplitude (see Eq.~(\ref{eq:defrhov})), we have $\rho_V(\Xi^- \to \Lambda) < \rho_V(\Lambda \to p) < \rho_V(n \to p) <1$.  In contrast, $\rho_A$ is given by the superposition of the scalar density of baryon transition amplitude (see Eq.~(\ref{eq:defrhoa})), and thus we obtain $\rho_A(n \to p) < \rho_A(\Lambda \to p) < \rho_A(\Xi^- \to \Lambda) < 1$.  In the case of $f_1$, because the two SU(3) breaking effects from $f_1^{(0)}$ and $\rho_V$ work in {\em opposite} direction, we have $\delta f_1 < \delta f_1^{(0)}$. In the case of $g_1$, the situation is similar and the cm correction tends to compensate for part of the relativistic reduction effect, that is, we find $\delta g_1 < \delta g_1^{(0)}$.  
We note that, because we have used the constituent quark model, the obtained $\delta g_1$ is, as expected, smaller than that in the bag model~\cite{chodos74bag, cbm81, tonyadv13}. 

In vacuum we obtain that $\delta f_1(n \to p) \sim {\cal O}(10^{-6}) \ll \delta f_1(\Xi^- \to \Lambda) <  \delta f_1(\Lambda \to p) \sim {\cal O}(10^{-2})$ and $\delta g_1(\Lambda \to p) \sim {\cal O}(10^{-2}) < \delta g_1(\Xi^- \to \Lambda) < \delta g_1(n \to p) \sim {\cal O}(10^{-1})$.  This is consistent with the results in Refs.~\cite{donoghue82quark, eich85static}.

\newpage

\begin{table*}[h!]
  \caption{\label{tab:weakcc-vacuum}
    Weak coupling constants in vacuum.
  }
  \begin{ruledtabular}
    \begin{tabular}{ccccccc}
    \multicolumn{7}{c}{$n\rightarrow p$} \\
    \hline
      case & $\delta f_1^{(0)} \times 10^6$ & $\rho_{V}$ & $\delta f_1 \times 10^6$ & $\delta g_1^{(0)}$ & $\rho_{A}$  & $\delta g_1$ \\
    \cline{1-1}     \cline{2-4}  \cline{5-7}
      1 & 3.12 & 0.999999789 & 2.91 & 0.187 & 0.9177 & 0.114 \\
      2 & 2.28 & 0.999999810 & 2.08 & 0.156 & 0.9195 & 0.082 \\
      3 & 1.71 & 0.999999825 & 1.54 & 0.131 & 0.9209 & 0.057 \\
      \hline \hline
    \multicolumn{7}{c}{$\Lambda\rightarrow p$} \\
    \hline
      case & $\delta f_1^{(0)}$ & $\rho_{V}$ & $\delta f_1$ & $\delta g_1^{(0)}$ & $\rho_{A}$ & $\delta g_1$ \\
      \cline{1-1}     \cline{2-4}  \cline{5-7}
      1 & 0.012 & 0.9988 & 0.011 & 0.154 & 0.9254 & 0.086 \\
      2 & 0.010 & 0.9989 & 0.009 & 0.131 & 0.9271 & 0.062 \\
      3 & 0.008 & 0.9990 & 0.007 & 0.111 & 0.9285 & 0.043 \\
       \hline \hline
    \multicolumn{7}{c}{$\Xi^{-}\rightarrow\Lambda$} \\
    \hline
      case & $\delta f_1^{(0)}$ & $\rho_{V}$ & $\delta f_1$ & $\delta g_1^{(0)}$ & $\rho_{A}$ & $\delta g_1$ \\
      \cline{1-1}     \cline{2-4}  \cline{5-7}
      1 & 0.012 & 0.9973 & 0.009 & 0.154 & 0.9404 & 0.101 \\
      2 & 0.010 & 0.9975 & 0.007 & 0.131 & 0.9420 & 0.077 \\
      3 & 0.008 & 0.9976 & 0.006 & 0.111 & 0.9433 & 0.058 \\
    \end{tabular}
  \end{ruledtabular}
\end{table*}
\clearpage

\section{Quark-meson coupling model for asymmetric nuclear matter} \label{sec:qmcmodel}

In this section, we introduce the quark-meson coupling (QMC) model~\cite{guichon88a, saito94a, saito07nucleon, guichon18qmc, krein18nucl}, which starts with confined quarks as the degrees of freedom, with the relativistic confinement potential of the scalar-vector HO type (see Eq.~(\ref{eq:conpot})).  We assume that the strength parameter $c$ in the potential does not change in matter. 

We here consider the mean fields of $\sigma$, $\omega$, $\rho$ and $\delta$ mesons, which interact with the confined quarks, in uniformly distributed, asymmetric nuclear matter. 
Let the mean-field values for the $\sigma$, $\omega$ (the time component), $\rho$ (the time component in the 3rd direction of isospin) and $\delta$ (in the 3rd direction of isospin) fields be ${\bar \sigma}$, ${\bar \omega}$, ${\bar \rho}$ and ${\bar \delta}$, respectively.  The Dirac equation for the quark field $\psi_{i}$ ($i = u$, $d$ or $s$) is then given by~\cite{saito94ns}
\begin{equation}
  \left[i\gamma \cdot \partial -\left(m_{i}-V_{s}\right)-\gamma_{0}V_{0}-\frac{c}{2}\left( 1+\gamma_{0}\right) r^2 \right]\psi_{i}({\vec r}, t)=0 , 
  \label{eq:Dirac-eq1}
\end{equation}
where $V_{s} = g^{q}_{\sigma}\bar{\sigma} + \tau_{3} g^{q}_{\delta} \bar{\delta}$ and
$V_{0} = g^{q}_{\omega}\bar{\omega} + \tau_{3} g^{q}_{\rho} \bar{\rho}$ ($\tau_3 = \pm 1$ for $\binom{u}{d}$ quark) with the quark-meson coupling constants, $g^{q}_{\sigma}$, $g^{q}_{\delta}$, $g^{q}_{\omega}$ and $g^{q}_{\rho}$.  Motivated by the Okubo-Zweig-Iizuka (OZI) rule, we here assume that the $\sigma$, $\omega$, $\rho$ and $\delta$ mesons couple to the $u$ and $d$ quarks only, not to the $s$ quark.  This breaks SU(3) symmetry explicitly.  Furthermore, the isoscalar $\sigma$ meson couples to the $u$ and $d$ quarks {\em equally}, while the isovector $\delta$ meson couples to the $u$ and $d$ quarks {\em oppositely}.  This breaks isospin symmetry. Now we respectively define the effective quark mass and the effective single-particle energy as $m_{i}^{\ast} \equiv m_{i}-V_{s} = m_i - g^{q}_{\sigma}\bar{\sigma} \mp g^{q}_{\delta} \bar{\delta}$ and $\epsilon_i^{\ast} \equiv \epsilon_{i}-V_{0} = \epsilon_i - g^{q}_{\omega}\bar{\omega} \mp g^{q}_{\rho} \bar{\rho}$ for $\binom{u}{d}$ quark, where $\epsilon_{i}$ is the eigenenergy of Eq.~(\ref{eq:Dirac-eq1}).  Note that $m_s^\ast = m_s$ and $\epsilon_s^\ast = \epsilon_s$. 

The static, lowest-state wave function in matter is presented by
\begin{equation}
  \psi_{i}({\vec r}, t)
  = \exp\left[ -i\epsilon_{i}t \right] \psi_i({\vec r}\,)  .
  \label{eq:sol1}
\end{equation}
The wave function $\psi_i({\vec r}\,)$ is then given by Eqs.~(\ref{eq:sol}) and (\ref{eq:norm}), in which $\epsilon_i$, $m_{i}$, $\lambda_{i}$ and $a_{i}$ are replaced with $\epsilon_i^{\ast}$, $m_{i}^{\ast}$, $\lambda_{i}^{\ast}$ and $a_{i}^{\ast}$, and the effective energy $\epsilon_i^{\ast}$ is determined by $\sqrt{\epsilon_i^\ast + m_i^\ast} (\epsilon_i^\ast - m_i^\ast) = 3 \sqrt{c}$ (see Eq.~(\ref{eq:qenergy})).   

The zeroth-order energy $E_B^{0\ast}$ of the low-lying baryon is thus given by 
\begin{equation}
E_B^{0 \ast} = \sum_i \epsilon_i^\ast   , 
  \label{eq:energy1}
\end{equation}
and we have the effective baryon mass in matter   
\begin{equation}
  M_B^\ast = E_B^{0 \ast} + E^{spin}_B - E^{cm \ast}_B . 
  \label{eq:bmass1}
\end{equation}
Here, we assume that the spin correlation $E^{spin}_B$ does not change in matter, and the cm correction $E^{cm \ast}_B$ is given by Eqs.~(\ref{eq:cmcorrection1}) -  (\ref{eq:cmcorrection4}) with $\epsilon_i^{\ast}$ and $m_{i}^{\ast}$, instead of $\epsilon_i$ and $m_{i}$.  

For describing asymmetric nuclear matter, we now start from the following Lagrangian density in mean-field approximation 
\begin{align}
  {\cal L} & = {\bar \psi}_N \left[ i \gamma \cdot \partial - M^\ast_N({\bar \sigma}, {\bar \delta}) - g_\omega \gamma_0 {\bar \omega} - g_\rho \gamma_0 \tau_3 {\bar \rho} \right] \psi_N  \nonumber \\
  & - \frac{m_\sigma^2}{2} {\bar \sigma}^2 - \frac{m_\delta^2}{2} {\bar \delta}^2 + \frac{m_\omega^2}{2} {\bar \omega}^2 + \frac{m_\rho^2}{2} {\bar \rho}^2 - \frac{g_2}{3} {\bar \sigma}^3 
  \label{eq:lagrangian}
\end{align}
with $\psi_N$ the (isodoublet) nucleon field, $\tau_3$ the 3rd component of Pauli matrix and 
$M^\ast_N({\bar \sigma}, {\bar \delta})$ the mass matrix whose component is given by Eq.~(\ref{eq:bmass1}).  The nucleon-meson coupling constants, $g_\sigma$,  $g_\omega$, $g_\rho$ and $g_\delta$, are respectively related to the quark-meson coupling constants as $g_\sigma = 3g_\sigma^q$,  $g_\omega = 3g_\omega^q$, $g_\rho = g_\rho^q$ and $g_\delta = g_\delta^q$.  The meson masses are taken to be $m_\sigma = 550$ MeV, $m_\omega = 783$ MeV, $m_\rho = 770$ MeV and $m_\delta = 983$ MeV.  We add the last term to the Lagrangian, which is the nonlinear, self-coupling term of $\sigma$ meson, in order to reproduce the properties of nuclear matter as discussed later. Here, we do not include the nonlinear term $\frac{1}{4} g_3 \sigma^4$, because it plays similar roles as $\frac{1}{3} g_2 \sigma^3$.  

The total energy per nucleon of asymmetric nuclear matter is then obtained by 
\begin{align}
  E
  &= \sum_{j=p, n} \frac{1}{\pi^{2}\rho_N}\int_{0}^{k_{F_{j}}} dk \,k^{2} \sqrt{M_{j}^{\ast 2}+k^{2}}
    + g_{\omega}\bar{\omega} + g_{\rho} \left( \frac{\rho_3}{\rho_N} \right) \bar{\rho} \nonumber \\
  &+ \frac{1}{2\rho_N}\left(m_{\sigma}^{2}\bar{\sigma}^{2}-m_{\omega}^{2}\bar{\omega}^{2}+m_{\delta}^{2}\bar{\delta}^{2}-m_{\rho}^{2}\bar{\rho}^{2}\right)
    +  \frac{1}{3\rho_N}g_{2}\bar{\sigma}^{3} , 
    \label{eq:energydensity}
\end{align}
where $k_{F_{p(n)}}$ is the Fermi momentum for protons (neutrons), and this is related to the density of protons (neutrons), $\rho_{p(n)}$, through $\rho_{p(n)} = k_{F_{p(n)}}^3/(3\pi^2)$. 
The total nucleon density is given by $\rho_N = \rho_p + \rho_n$, and the difference in proton and neutron densities is defined by $\rho_3 \equiv \rho_p - \rho_n$.  Using Eq.~(\ref{eq:energydensity}), we can calculate pressure by $P = \rho_N^2 \left( \partial E/\partial \rho_N \right)$.  Then, the binding energy per nucleon, $E_b$, is defined by 
\begin{equation}
  E_{b}(\rho_{N},\alpha)
  \equiv E(\rho_{N},\alpha) -\frac{1}{2}\left[(M_{n}+M_{p})+(M_{n}-M_{p})\alpha\right]
    \label{eq:binding-energy}
\end{equation}
where $\alpha=(\rho_{n}-\rho_{p})/\rho_{N}$ and $E(\rho_{N},\alpha)$ is given by Eq.~(\ref{eq:energydensity}).   For detail, see Appendix~\ref{app:asymprop}. 

The mean-field values for the mesons satisfy~\cite{saito94ns} 
\begin{align}
  & \left(m_{\sigma}^{2}+g_{2}\bar{\sigma} \right)\bar{\sigma}
  = - \sum_{j=p, n} \left( \frac{\partial M_j^\ast}{\partial {\bar \sigma}} \right) \rho_{j}^{s} 
  \equiv g_\sigma \sum_{j} G_{\sigma j}(\bar{\sigma},\bar{\delta}) \rho_{j}^{s}  , 
  \label{eq:sigmamf} \\
  & m_{\omega}^{2} \bar{\omega} = g_{\omega}\rho_{N}, \\
  & m_{\delta}^{2}\bar{\delta} 
  = - \sum_{j} \left( \frac{\partial M_j^\ast}{\partial {\bar \delta}} \right) \rho_{j}^{s} 
  \equiv g_\delta \sum_{j} G_{\delta j}(\bar{\sigma},\bar{\delta}) \rho_{j}^{s}  , \\
  & m_{\rho}^{2}\bar{\rho} = g_{\rho} \rho_{3},  \label{eq:rhomf}
\end{align}
with $G_{\sigma j}$ ($G_{\delta j}$) the isoscalar (isovector) quark scalar density in $j$ ($=$ proton or neutron) (see Appendix~\ref{app:qscalardensity}), and $\rho_{j}^{s}$ the scalar density of $j$ in matter 
\begin{equation}
 \rho_{j}^{s} = \frac{1}{\pi^{2}} \int_{0}^{k_{F_{j}}} dk \,k^{2} \frac{M_{j}^{\ast}}{\sqrt{M_{j}^{\ast 2}+k^{2}}}  . 
  \label{eq:scalardenj}
\end{equation}

Now we have five parameters, $g_\sigma$, $g_\omega$, $g_\rho$, $g_\delta$ and $g_2$ in the present QMC model.  
The coupling constants, $g_\sigma$ and $g_\omega$, are determined so as to fit the binding energy, $E_b(\rho_0, 0) = -16$ MeV, at the saturation density, $\rho_0 = 0.15$ fm$^{-3}$, for equilibrium, symmetric nuclear matter. The $\delta$-nucleon coupling constant is adopted to be $g_\delta^2/4\pi = 1.30$, $2.49$ or $4.72$~\cite{mach89delta}. 
We fit the $\rho$-nucleon coupling constant so as to reproduce the bulk symmetry energy of symmetric nuclear matter, $E_{sym}(\rho_0) = 32.0$ MeV.  Finally, the nonlinear coupling constant, $g_2$, is used to fit the nuclear incompressibility, $K_0 = 240$ MeV.  

\begin{figure*}[h!]
  \includegraphics[width=8.0cm,keepaspectratio,clip]{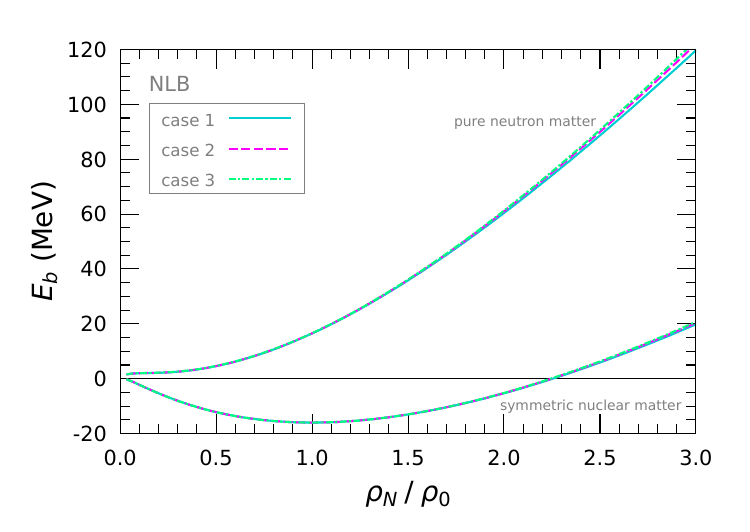}%
  \includegraphics[width=8.0cm,keepaspectratio,clip]{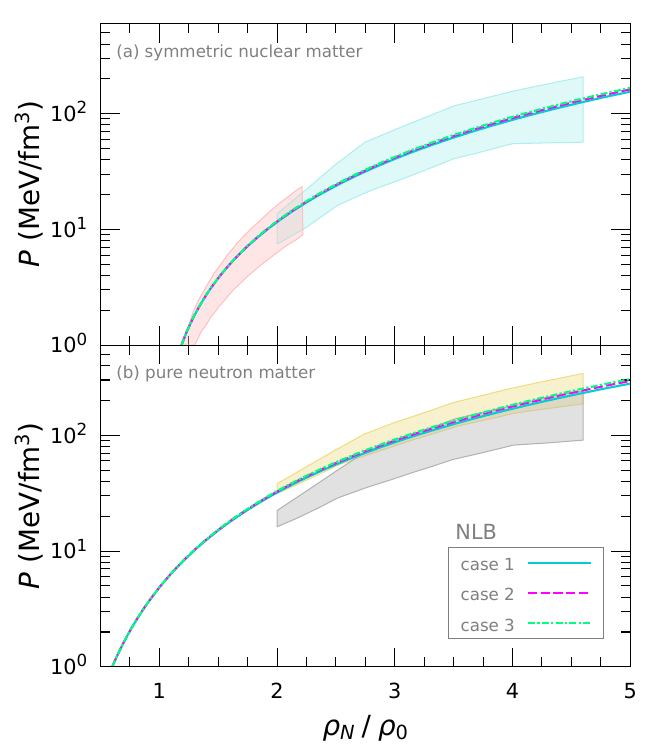}%
  \caption{\label{fig:Eb-Press-NLB}
   Binding energy per nucleon (left panel) and pressure (right panel) as a function of $\rho_N/\rho_0$.  The calculation is performed with the NLB model.  
   In the right panel, the shaded regions are given by the analyses of particle flow data in heavy ion collisions (see Refs.~\cite{Danielewicz:2002pu, Fuchs:2005zg, Lynch:2009vc}). 
   }
\end{figure*}

The coupling constants and the properties of nucleons and nuclear matter are respectively listed in Tables~\ref{tab:CCs} and \ref{tab:matter-properties}.  Note that, in the model NL0, the $\delta$ meson is not included, while, in the model NLA, NLB or NLC, $g_\delta^2/4\pi$ is taken to be $1.30$, $2.49$ or $4.72$, respectively.  At the saturation density, the root-mean-square charge radius of proton increases by about $5\%$, compared with the radius in vacuum, which may be consistent with the experiments~\cite{lu98in}.  We also calculate the slope and the curvature parameters of nuclear symmetry energy, $L$ and $K_{sym}$, (see Appendix~\ref{app:asymprop}), which are related to astrophysical multi-messenger observations of neutron stars~\cite{miyatsu22asy, miyatsu23can}.  The binding energy per nucleon and pressure are illustrated in Fig.~\ref{fig:Eb-Press-NLB}. 

\begin{figure*}[h!]
  \includegraphics[width=8.0cm,keepaspectratio,clip,page=1]{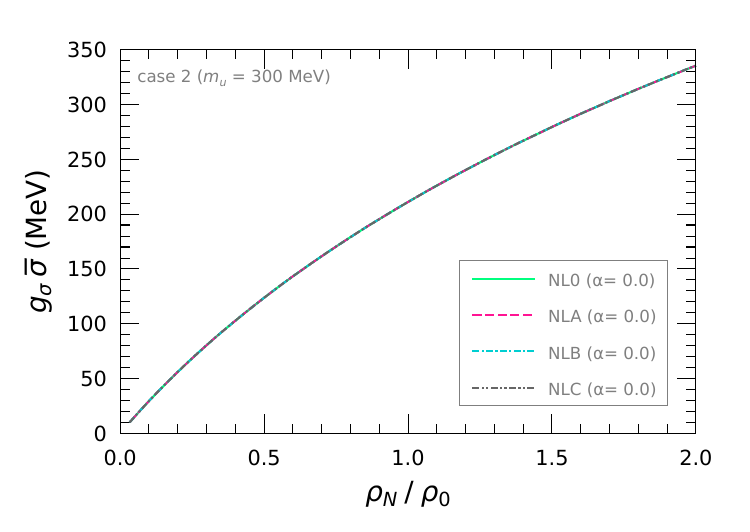}%
  \includegraphics[width=8.0cm,keepaspectratio,clip,page=1]{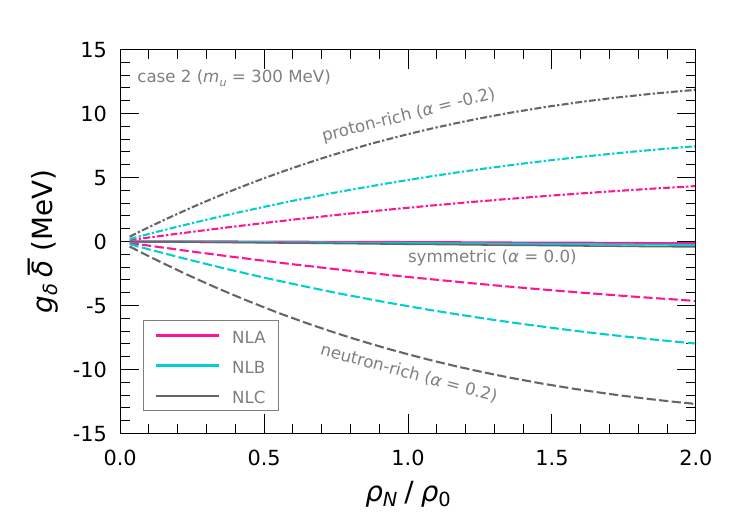}%
  \caption{\label{fig:mean-fields}
    Mean-field values of the meson fields as a function of $\rho_N/\rho_0$. 
    We show the results in case 2 only in Table~\ref{tab:mean-fields}.  
    The left panel is for the $\sigma$ field in symmetric nuclear matter.  The model dependence of the $\sigma$ field is very small.  The right panel is for the $\delta$ field in case of $\alpha = -0.2$ (the dot-dashed curves), $0$ (the solid curves) or $0.2$ (the dashed curves).  Note that $\alpha = 0.2$ corresponds to ${}^{208}_{\phantom{1}82}$Pb. 
   }
\end{figure*}

In Table~\ref{tab:mean-fields}, we show the mean-field values for the mesons in symmetric nuclear matter.  In this case, the $\rho$ meson field vanishes.  In contrast, the $\delta$ meson field has a negative, small value even in symmetric nuclear matter, which is caused by the $u$-$d$ quark mass difference~\cite{saito94ns}.   Furthermore, the density dependence of the mean-fields is depicted in Fig.~\ref{fig:mean-fields}.  

\begin{figure*}[h!]
  \includegraphics[width=8.0cm,keepaspectratio,clip,page=1]{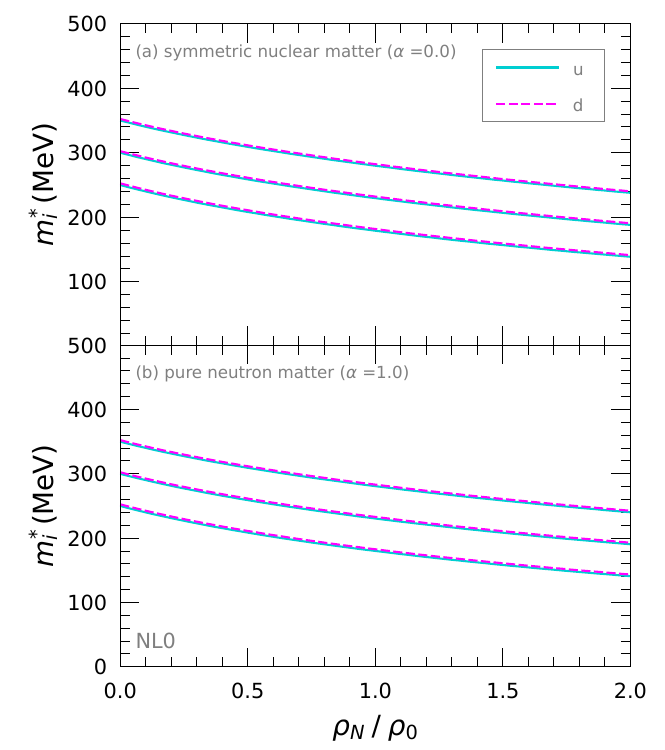}%
  \includegraphics[width=8.0cm,keepaspectratio,clip,page=1]{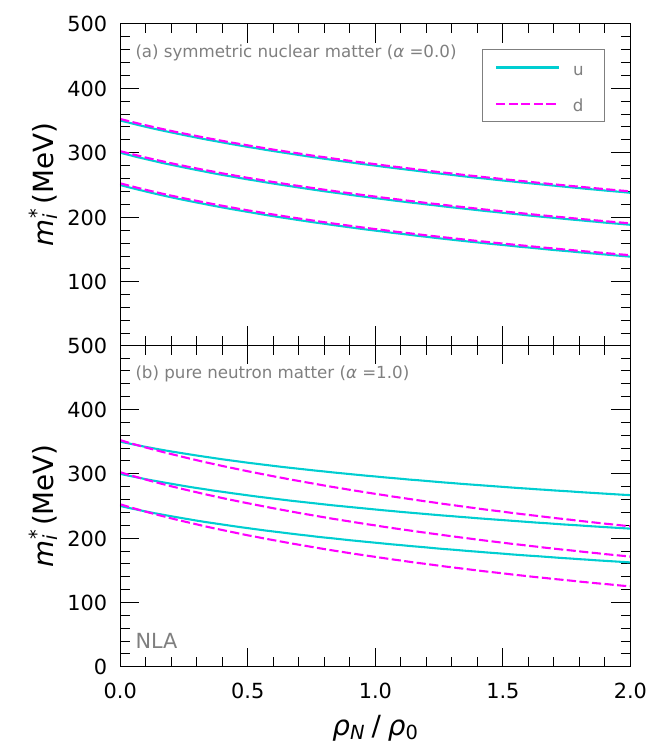} \\%
  \includegraphics[width=8.0cm,keepaspectratio,clip,page=1]{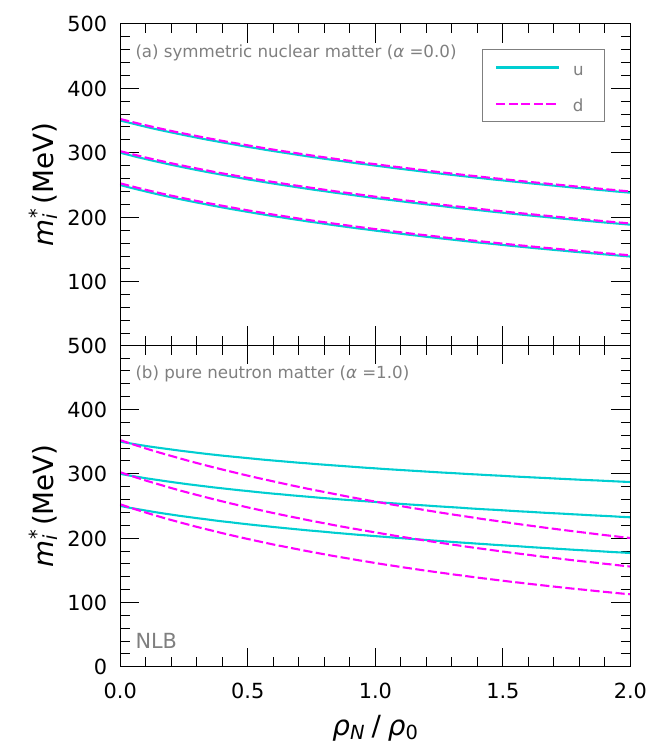}%
  \includegraphics[width=8.0cm,keepaspectratio,clip,page=1]{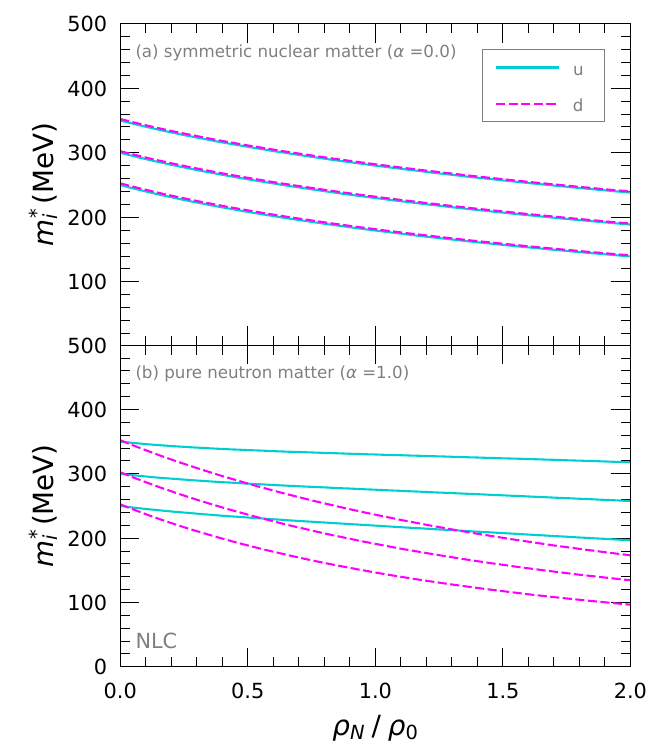}%
  \caption{\label{fig:Quark-mass}
   Effective quark mass ($i = u$ or $d$) in the NL0, NLA, NLB or NLC model.  In each panel, the bottom (middle) [top] two curves are for case 1 (2) [3]. 
   }
\end{figure*}

\begin{figure*}[h!]
  \includegraphics[width=8.0cm,keepaspectratio,clip,page=1]{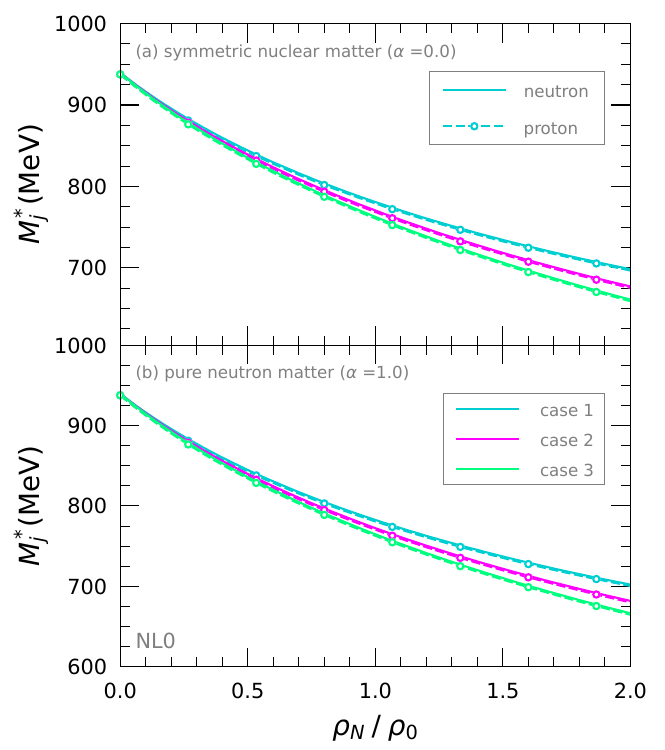}%
  \includegraphics[width=8.0cm,keepaspectratio,clip,page=1]{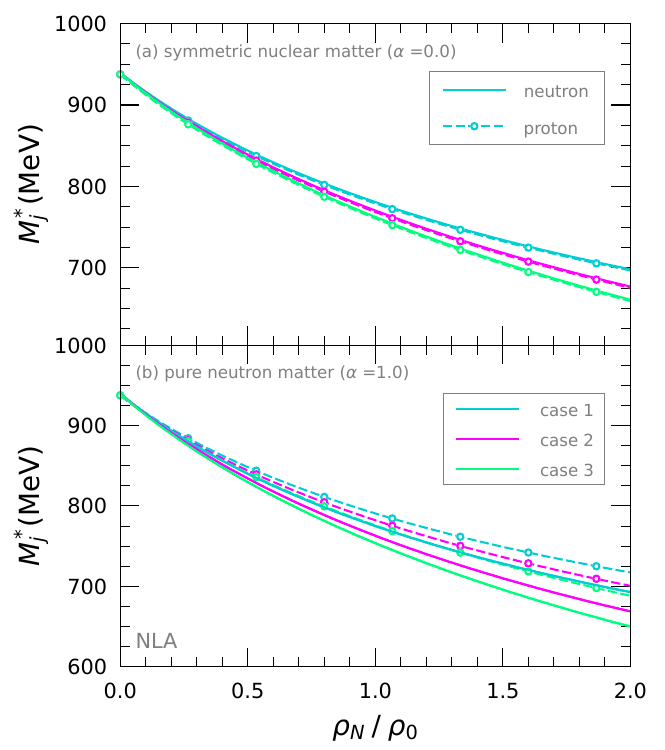} \\ %
  \includegraphics[width=8.0cm,keepaspectratio,clip,page=1]{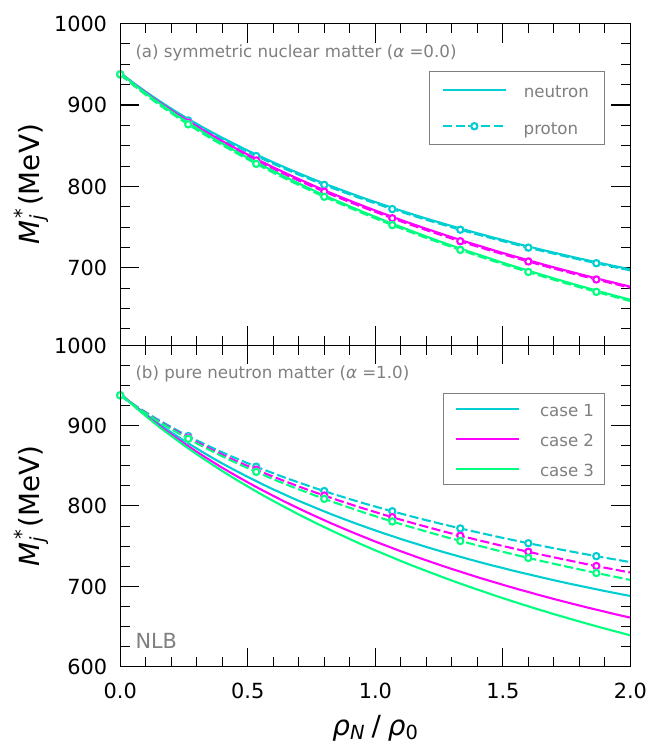}%
  \includegraphics[width=8.0cm,keepaspectratio,clip,page=1]{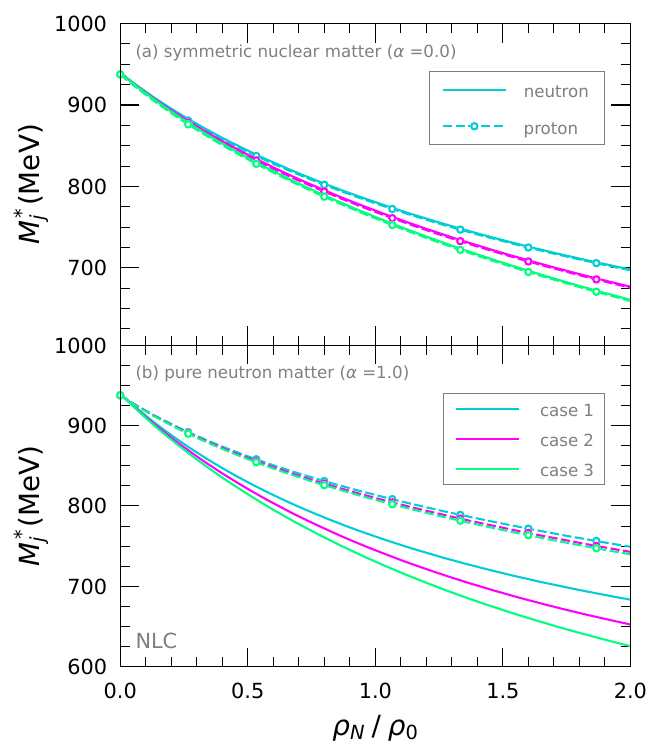}%
  \caption{\label{fig:Nucleon-mass}
   Effective nucleon mass ($j = p$ or $n$) in the NL0, NLA, NLB or NLC model.  
   }
\end{figure*}

In Figs.~\ref{fig:Quark-mass} and \ref{fig:Nucleon-mass}, we respectively show the density dependence of the effective quark mass and that of the effective nucleon mass in symmetric nuclear matter or pure neutron matter.  At high density the quark mass in pure neutron matter is considerably modified by the $\delta$ field, in which the $d$ quark mass is much lighter than the $u$ quark mass. 
This tendency also appears in the effective nucleon mass, that is, at high density the neutron mass decreases very rapidly in pure neutron matter, compared with the proton mass. 
In contrast, the $\delta$ field affects the mass little in symmetric nuclear matter.  

\begin{figure*}[h!]
  \includegraphics[width=8.0cm,keepaspectratio,clip,page=1]{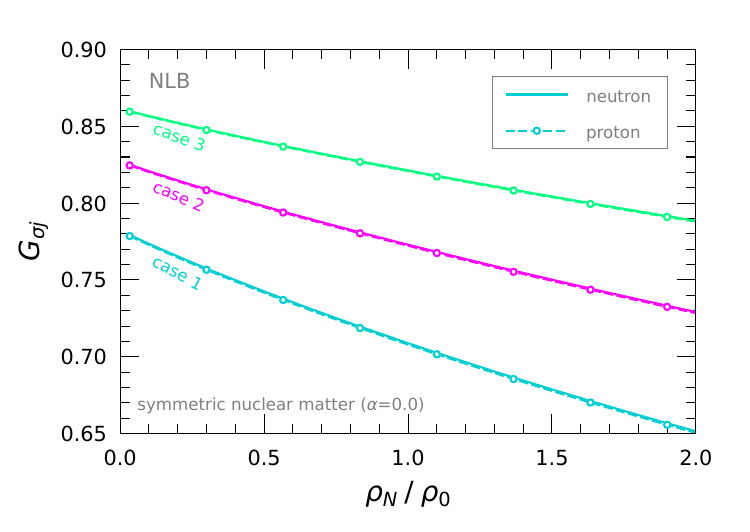}%
\\ %
  \includegraphics[width=8.0cm,keepaspectratio,clip,page=1]{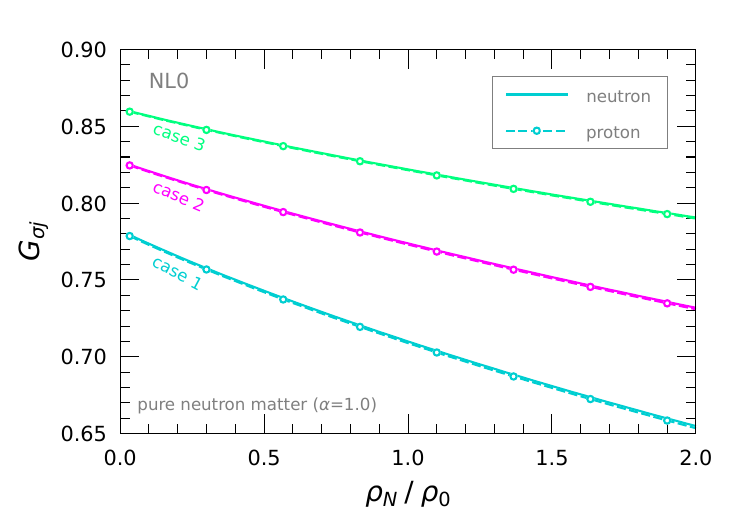}%
  \includegraphics[width=8.0cm,keepaspectratio,clip,page=1]{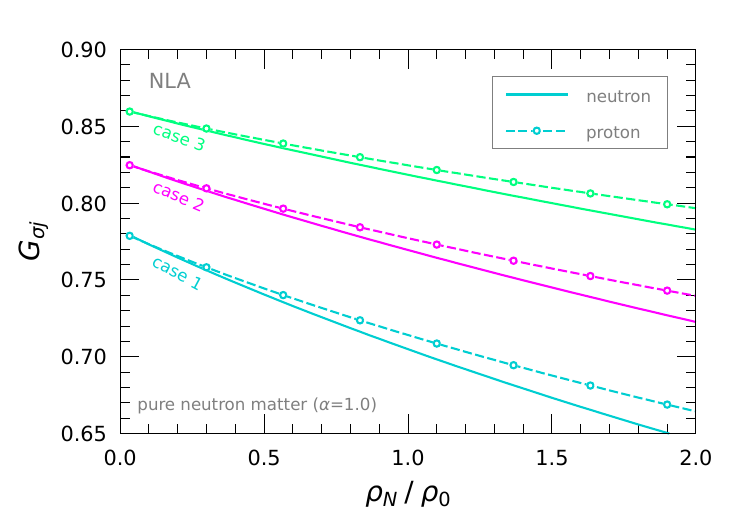} \\ %
  \includegraphics[width=8.0cm,keepaspectratio,clip,page=1]{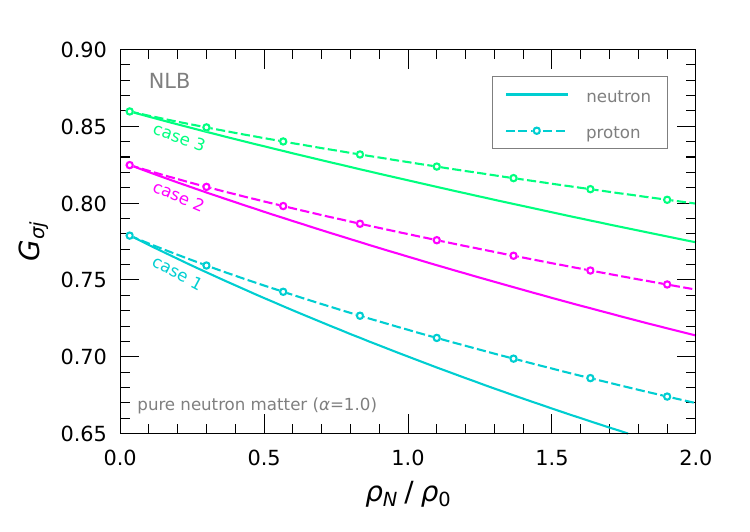}%
  \includegraphics[width=8.0cm,keepaspectratio,clip,page=1]{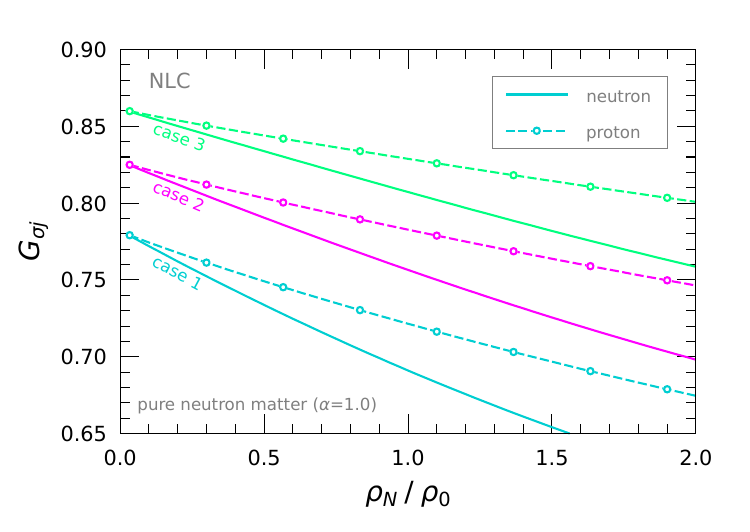}%
  \caption{\label{fig:Gsigma}
   Isoscalar, quark scalar density, $G_{\sigma j}$, as a function of $\rho_N/\rho_0$ ($j = p$ or $n$). 
   For symmetric nuclear matter (top panel), we show the result in the NLB model only, because the model dependence of $G_{\sigma j}$ is very small.  The lower four panels are for pure neutron matter. }
\end{figure*}

In Fig.~\ref{fig:Gsigma}, we illustrate the isoscalar, quark scalar densities, $G_{\sigma j}$, in proton or neutron.  In symmetric nuclear matter, because the absolute value of $\delta$ field is quite small (see Fig.~\ref{fig:mean-fields} and Table~\ref{tab:mean-fields}), the $u$ quark mass is slightly lighter , i.e.  {\em more relativistic}, than the $d$ quark mass (see Fig.~\ref{fig:Quark-mass}).  Therefore, we find that $G_{\sigma n}$ is slightly larger than $G_{\sigma p}$.  In contrast, in pure neutron matter, because the $\delta$ field is {\em negative} and its absolute value is larger with increasing the nuclear density, the $d$ quark mass is lighter ({\em more relativistic}) than the $u$ quark mass (see Fig.~\ref{fig:Quark-mass}).  Thus, we obtain $G_{\sigma p} > G_{\sigma n}$ in neutron matter. 

\begin{figure*}[h!]
  \includegraphics[width=8.0cm,keepaspectratio,clip,page=1]{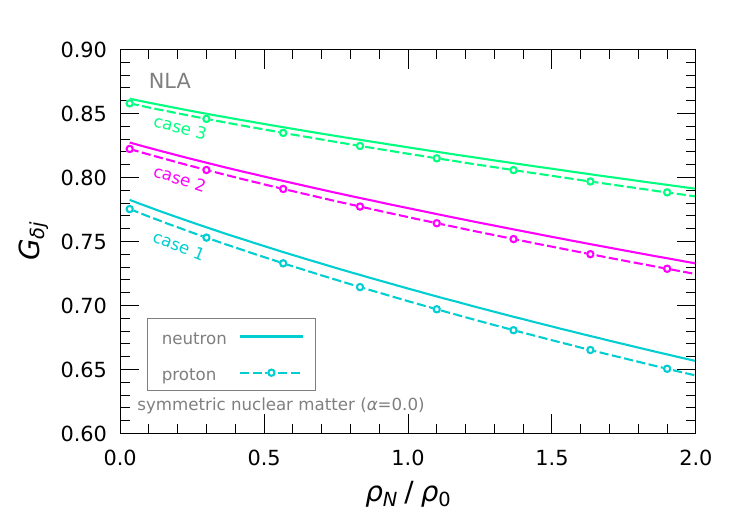}%
  \includegraphics[width=8.0cm,keepaspectratio,clip,page=1]{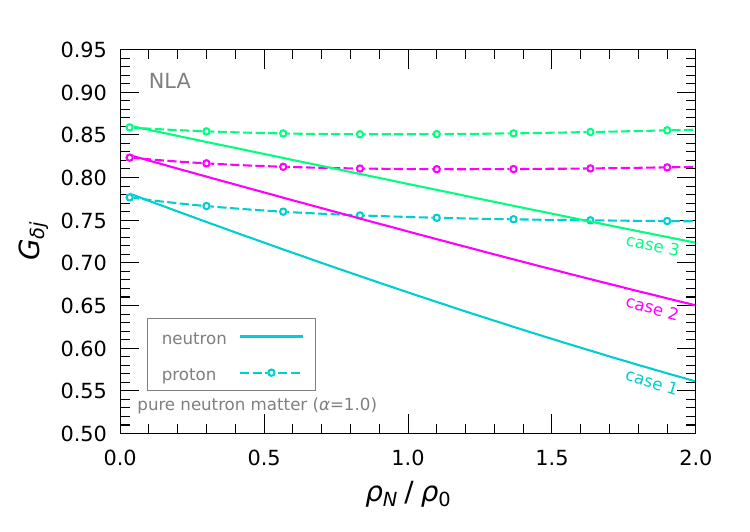} \\ %
  \includegraphics[width=8.0cm,keepaspectratio,clip,page=1]{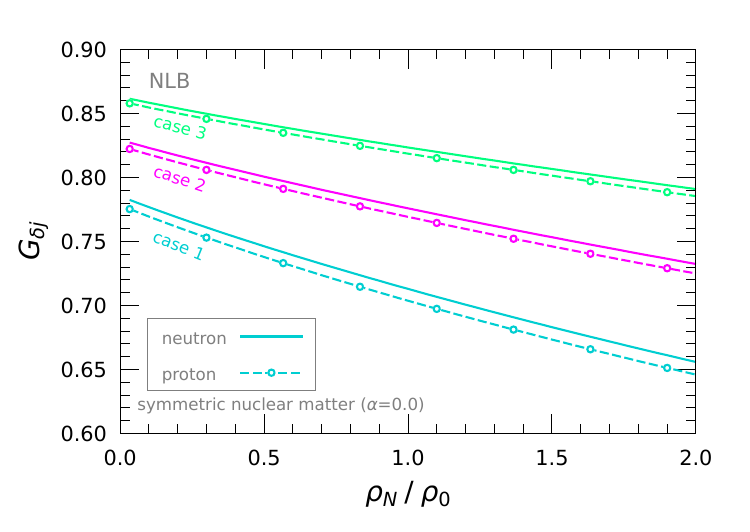}%
  \includegraphics[width=8.0cm,keepaspectratio,clip,page=1]{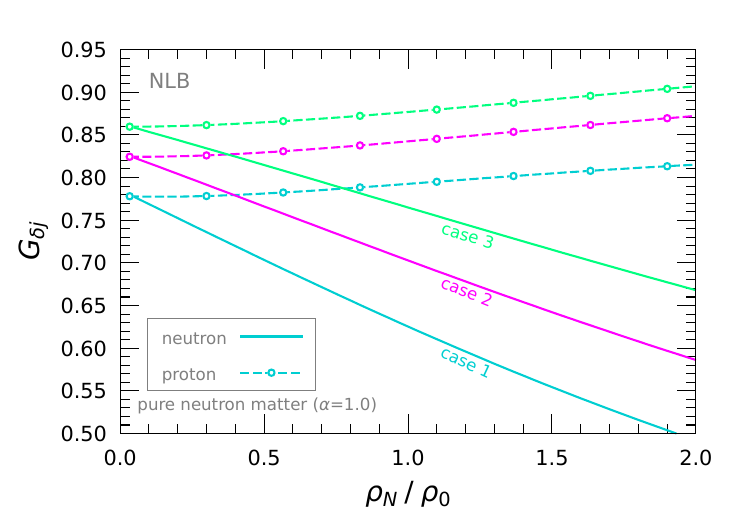} \\ %
  \includegraphics[width=8.0cm,keepaspectratio,clip,page=1]{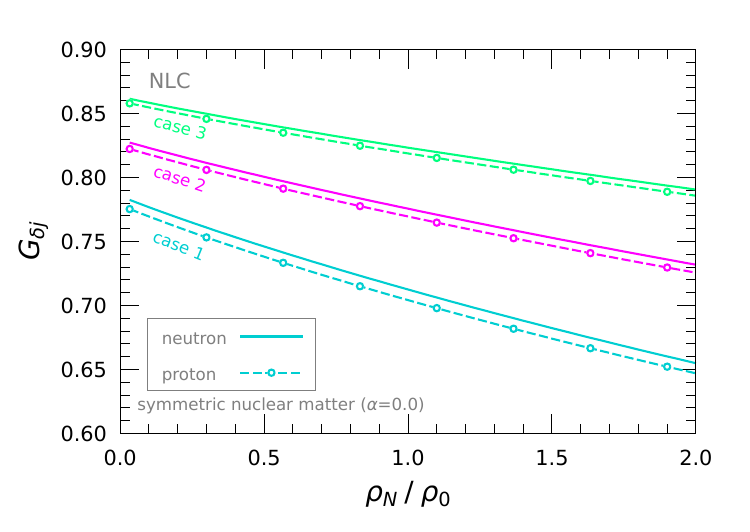}%
  \includegraphics[width=8.0cm,keepaspectratio,clip,page=1]{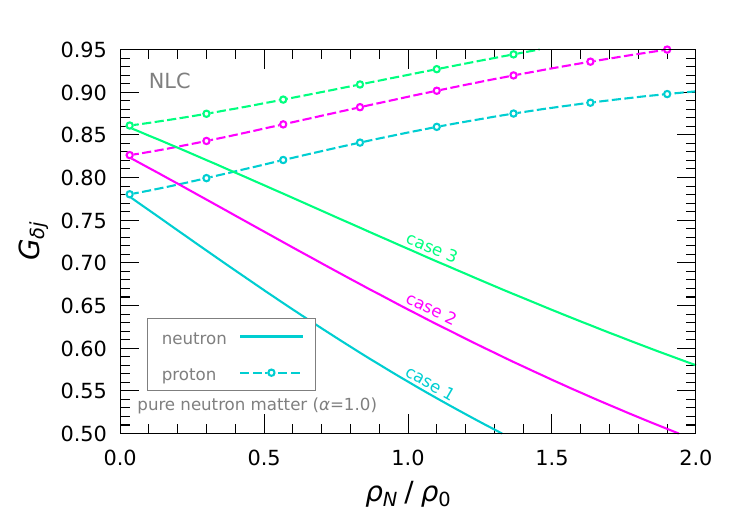}%
  \caption{\label{fig:Gdelta}
   Isovector, quark scalar density, $G_{\delta j}$, as a function of $\rho_N/\rho_0$ ($j = p$ or $n$).  We show the results in the NLA, NLB or NLC model. Note that $G_{\delta j} = 0$ in the NL0 model.  The left (right) three panels are for symmetric nuclear matter (pure neutron matter). 
   }
\end{figure*}

In Fig.~\ref{fig:Gdelta}, we show the isovector, quark scalar densities, $G_{\delta j}$.  Here  $G_{\delta n}$ is negative (see Eq.~(\ref{eq:derivdelta})), and thus we plot $|G_{\delta n}|$ in the figure.  In symmetric nuclear matter, because the $u$ quark is more relativistic than the $d$ quark as discussed above, we find that $|G_{\delta n}| > G_{\delta p}$ and the effect of the $\delta$ field is small.  However, in pure neutron matter, the situation is opposite, and the $d$ quark becomes more relativistic than the $u$ quark. Therefore, $G_{\delta p}$ is much larger than $|G_{\delta n}|$ in neutron matter.  

\begin{figure*}[h!]
  \includegraphics[width=8.0cm,keepaspectratio,clip,page=1]{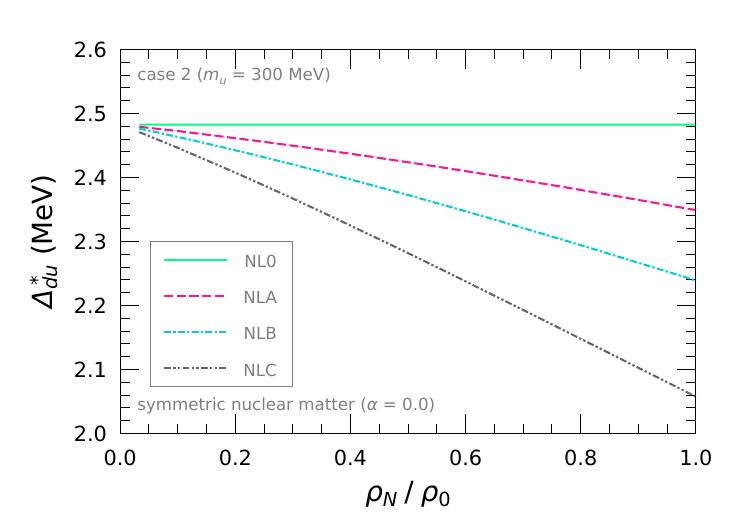} \\%
  \includegraphics[width=8.0cm,keepaspectratio,clip,page=1]{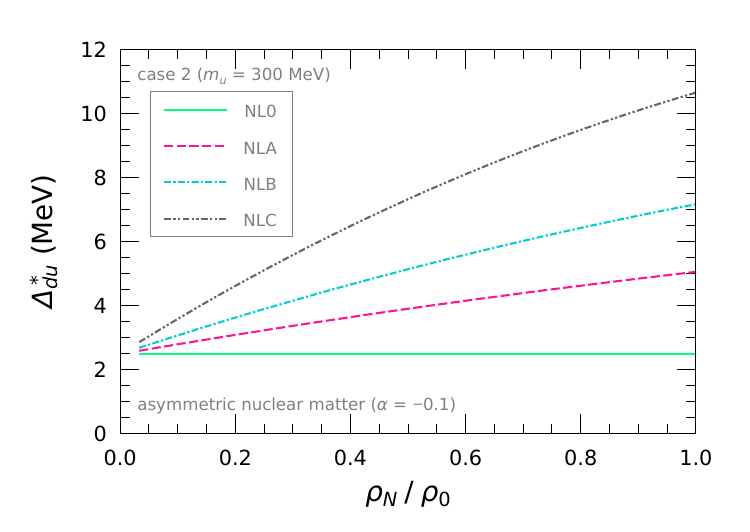}%
  \includegraphics[width=8.0cm,keepaspectratio,clip,page=1]{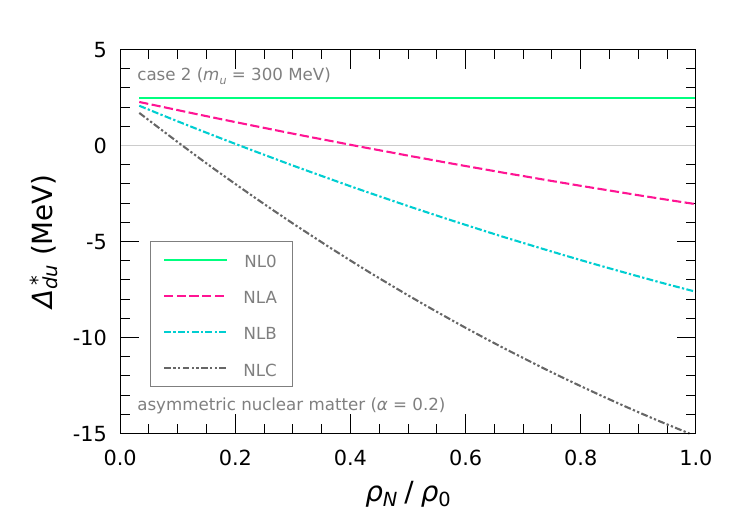}%
  \caption{\label{fig:Mass-diff}
   Quark mass difference $\Delta_{du}^\ast$ in matter. We show the results in case 2. In the NL0 model, $\Delta_{du}^\ast$ does not change in matter, i.e. $\Delta_{du}^\ast = \Delta_{du}$. The top panel is for symmetric nuclear matter, while the bottom two panels are for asymmetric nuclear matter, $\alpha = -0.1$ (left) and $0.2$ (right).  
   }
\end{figure*}

Now we emphasize the following interesting feature.  The effect of the isovector scalar field on the effective quark mass is small in symmetric nuclear matter, while its contribution to the mass is very significant in neutron-rich or in proton-rich matter.  In Fig.~\ref{fig:Mass-diff}, we illustrate the difference in the $u$ and $d$ quark masses  
\begin{equation}
 \Delta_{du}^\ast = m_d^\ast - m_u^\ast = \Delta_{du} + 2 g_\delta {\bar \delta}  , 
  \label{eq:dumassdiff}
\end{equation}
where, for example, $\Delta_{du} \sim 2.5$ MeV in case 2 (see Table~\ref{tab:parameters}). Note that the isoscalar $\sigma$ meson does not contribute to the mass difference $\Delta_{du}^\ast$.  Because the $\delta$ field takes a negative, small value in symmetric nuclear matter (see Fig.~\ref{fig:mean-fields} and Table~\ref{tab:mean-fields}), $\Delta_{du}^\ast$ slowly decreases with increasing $\rho_N$ (see the top panel in Fig.~\ref{fig:Mass-diff}).  Contrary to this behavior, 
in neutron-rich matter ($\alpha = 0.2$, which corresponds to ${}^{208}_{\phantom{1}82}$Pb), the mass difference reduces rapidly, and, at $\rho_N/\rho_0 \sim$ 0.4 (0.2) [0.1], $\Delta_{du}^\ast$ {\em vanishes} in the NLA (NLB) [NLC] model, which implies that {\em isospin symmetry is restored}.  Beyond that density, $\Delta_{du}^\ast$ becomes negative and the absolute value increases as the density grows up.  Thus, the $\delta$ field significantly enhances the breaking of isospin symmetry around $\rho_0$. 
In contrast, in proton-rich matter (for example, $\alpha = -0.1$), $\Delta_{du}^\ast$ is positive and increases monotonously, and thus the symmetry breaking is also enhanced.  This feature is robust and may not depend on the model we assume.  Furthermore, such behavior of the isovector scalar field may help to understand the Okamoto-Nolen-Schiffer anomaly for heavy nuclei~\cite{saito94ns}.

\newpage

\begin{table*}[h!]
  \caption{\label{tab:CCs}
    Coupling constants.  For detail, see the text. 
  }
  \begin{ruledtabular}
    \begin{tabular}{lcccccc}
      Model                   & case & $g_{\sigma}$ & $g_{\omega}$ & $g_{\delta}$ & $g_{\rho}$ & $g_{2}$ (fm$^{-1}$) \\
      \hline
      \multirow{3}{*}{NL0} &           1 &        10.34 &         7.34 &          --- &       4.37 &               23.61 \\
      \                        &           2 &         9.98 &         7.69 &          --- &       4.35 &               24.97 \\
      \                        &           3 &         9.72 &         7.95 &          --- &       4.33 &               25.67 \\
      \hline
      \multirow{3}{*}{NLA} &          1 &        10.34 &         7.34 &         4.04 &       4.83 &               23.61 \\
      \                        &           2 &         9.98 &         7.69 &         4.04 &       4.90 &               24.97 \\
      \                        &           3 &         9.72 &         7.95 &         4.04 &       4.96 &               25.67 \\
      \hline
      \multirow{3}{*}{NLB} &         1 &        10.34 &         7.34 &         5.59 &       5.16 &               23.61 \\
      \                        &           2 &         9.98 &         7.69 &         5.59 &       5.31 &               24.97 \\
      \                        &           3 &         9.72 &         7.95 &         5.59 &       5.43 &               25.67 \\
      \hline
      \multirow{3}{*}{NLC} &         1 &        10.34 &         7.34 &         7.70 &       5.65 &               23.61 \\
      \                        &           2 &         9.98 &         7.69 &         7.70 &       5.93 &               24.97 \\
      \                        &           3 &         9.72 &         7.95 &         7.70 &       6.16 &               25.67 \\
    \end{tabular}
  \end{ruledtabular}
\end{table*}
\begin{table*}[h!]
  \caption{\label{tab:matter-properties}
    Properties of in-medium nucleons and symmetric nuclear matter at the saturation density.
    All values except for the charge radii are given in MeV.  The values of $\braket{r^{2}}_{n}$ and $\braket{r^{2}}_{p}^{1/2}$ are in fm$^{2}$ and in fm, respectively.  The nuclear incompressibility and the symmetry energy are 
    respectively fixed as $K_0 = 240$ MeV and $E_{sym} = 32.0$ MeV.  Here, $L$ and $K_{sym}$ are respectively the slope and the curvature parameters of nuclear symmetry energy (see Appendix~\ref{app:asymprop}).    
  }
    \begin{ruledtabular}
    \begin{tabular}{lccccccccccr}
      Model                   & case & $M_{n}^{\ast}$ & $M_{p}^{\ast}$ & $E_{n}^{0\ast}$ & $E_{p}^{0\ast}$ & $E_{n}^{cm\ast}$ & $E_{p}^{cm\ast}$ & $\braket{r^{2}}_{n}$ & $\braket{r^{2}}_{p}^{1/2}$ &   $L$ &  $K_{sym}$ \\
      \hline
      \multirow{3}{*}{NL0} &     1 &         780.9 &         779.0 &         1337.1 &         1335.4 &           319.4 &           319.5  &    0.0011 &      0.888 &  86.2 &      $-19.0$ \\
      \                        &     2 &         770.5 &         768.6 &         1401.0 &         1399.2 &           296.0 &           296.2   &                         0.0010 &                     0.883 &  86.6 &      $-16.6$ \\
      \                        &     3 &         762.6 &         760.7 &         1478.5 &         1476.7 &           273.9 &           274.1  &                            0.0008 &                     0.879 &  87.0 &      $-14.6$ \\
      \hline
      \multirow{3}{*}{NLA} &    1 &         780.8 &         779.1 &         1337.0 &         1335.4 &           319.4 &           319.5   &                       0.0011 &                     0.888 &  90.7 &          2.9 \\
      \                        &     2 &         770.4 &         768.6 &         1400.9 &         1399.3 &           296.0 &           296.2   &                       0.0010 &                     0.883 &  91.0 &          6.1 \\
      \                        &     3 &         762.6 &         760.7 &         1478.5 &         1476.7 &           273.9 &           274.1   &                    0.0008 &                     0.879 &  91.1 &          7.8 \\
      \hline
      \multirow{3}{*}{NLB} &    1 &         780.8 &         779.2 &         1337.0 &         1335.5 &           319.4 &           319.5  &                   0.0010 &                     0.888  &  95.8 &         26.7 \\
      \                        &     2 &         770.4 &         768.7 &         1400.9 &         1399.3 &           296.0 &           296.2   &                   0.0009 &                     0.883 &  96.0 &         31.9 \\
      \                        &     3 &         762.6 &         760.7 &         1478.4 &         1476.7 &           273.9 &           274.1   &                     0.0008 &                     0.879 &  95.9 &         33.9 \\
      \hline
      \multirow{3}{*}{NLC} &   1 &         780.7 &         779.2 &         1336.9 &         1335.6 &           319.4 &           319.5   &                    0.0009 &                     0.888 & 106.9 &         71.1 \\
      \                        &    2 &         770.3 &         768.7 &         1400.8 &         1399.4 &           296.0 &           296.1  &                   0.0008 &                     0.883 & 107.4 &         84.7 \\
      \                        &     3 &         762.5 &         760.8 &         1478.4 &         1476.8 &           273.9 &           274.1  &                   0.0007 &                     0.879 & 106.9 &         90.2 \\
    \end{tabular}
  \end{ruledtabular}
\end{table*}
\begin{table*}[h!]
  \caption{\label{tab:mean-fields}
    Mean-field values of the meson fields (in MeV) at the saturation density.  The $\rho$ meson field vanishes in symmetric nuclear matter.  
  }
  \begin{ruledtabular}
    \begin{tabular}{lcccc}
      Model                   & case & $\bar{\sigma}$ & $\bar{\omega}$ & $\bar{\delta}$  \\
      \hline
      \multirow{3}{*}{NL0} &     1 &          20.54 &          13.80 &            ---    \\
      \                        &     2 &          21.15 &          14.46 &            ---      \\
      \                        &     3 &          21.61 &          14.94 &            ---      \\
      \hline
      \multirow{3}{*}{NLA} &    1 &          20.54 &          13.80 &       $-0.023$    \\
      \                        &     2 &          21.15 &          14.46 &       $-0.017$  \\
      \                        &     3 &          21.61 &          14.94 &       $-0.012$  \\
      \hline
      \multirow{3}{*}{NLB} &    1 &          20.54 &          13.80 &       $-0.030$ \\
      \                        &     2 &          21.15 &          14.46 &       $-0.022$ \\
      \                        &     3 &          21.61 &          14.94 &       $-0.016$  \\
      \hline
      \multirow{3}{*}{NLC} &     1 &          20.54 &          13.80 &       $-0.037$ \\
      \                        &     2 &          21.15 &          14.46 &       $-0.028$ \\
      \                        &     3 &          21.61 &          14.94 &       $-0.020$ \\
    \end{tabular}
  \end{ruledtabular}
\end{table*}

\clearpage

\section{Weak coupling constants in matter} \label{sec:weakmatter}

In this section, we consider the weak coupling constants in asymmetric nuclear matter. 
The vector and axial-vector coupling constants are calculated by Eqs.~(\ref{eq:corgva}), (\ref{eq:exrhov}) and (\ref{eq:exrhoa}), in which the quark mass and energy, $m_i$ and $\epsilon_i$, should be replaced with $m_i^\ast$ and $\epsilon_i^\ast$.  Furthermore, in the calculation of the cm correction to the quark currents, we also replace the baryon mass and energy, $M_{B}$ and $E_{B}$, by the effective values, $M_{B}^\ast$ and $E_{B}^\ast$, in matter.

\subsection{Vector coupling constant in matter} \label{subsec:vectormatter}

\begin{figure*}[h!]
  \includegraphics[width=12.0cm,keepaspectratio,clip,page=1]{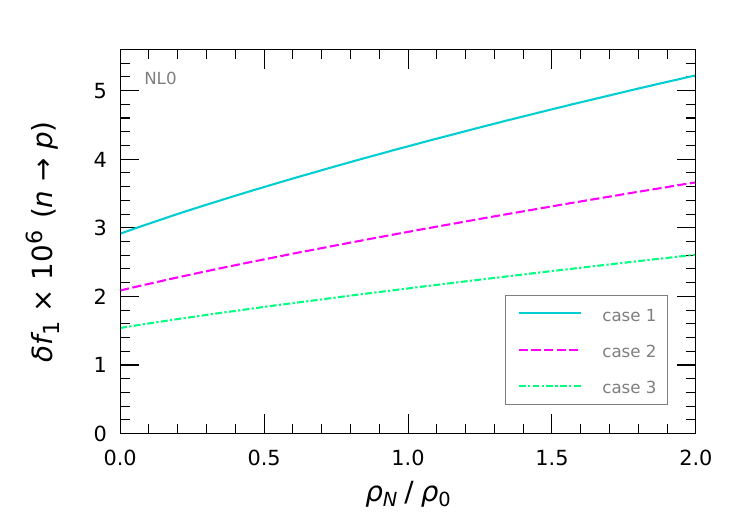}%
  \caption{\label{fig:FF-vector-NL0}
   Deviation, $\delta f_1$, for the neutron $\beta$ decay in the NL0 model with $\alpha = 0$.  The dependence of $\delta f_1$ on $\alpha$ is very small. 
   }
\end{figure*}

In Fig.~\ref{fig:FF-vector-NL0}, we first show the density dependence of $\delta f_1$ for the in-medium neutron $\beta$ decay in the NL0 model.  This is the case where the $\delta$ field is not included.  Then, $\delta f_1$ is of the order of $10^{-6}$, which is generated by the original $u$-$d$ quark mass difference only, namely $\Delta_{du}^\ast = \Delta_{du} \sim 2.5$ MeV (see Eq.~(\ref{eq:dumassdiff})).  We find that $\delta f_1$ gradually grows up with increasing the density, which is mainly due to the variation of $f_1^{(0)}$ in matter. 

\begin{figure*}[h!]
  \includegraphics[width=8.0cm,keepaspectratio,clip,page=1]{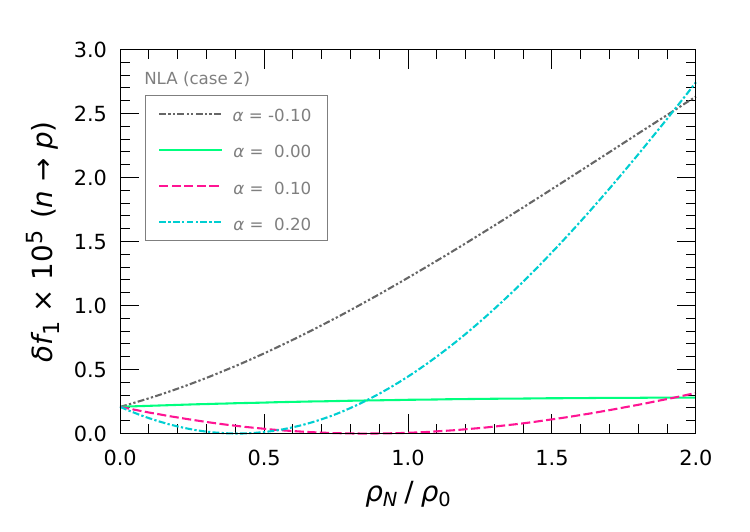}%
  \includegraphics[width=8.0cm,keepaspectratio,clip,page=1]{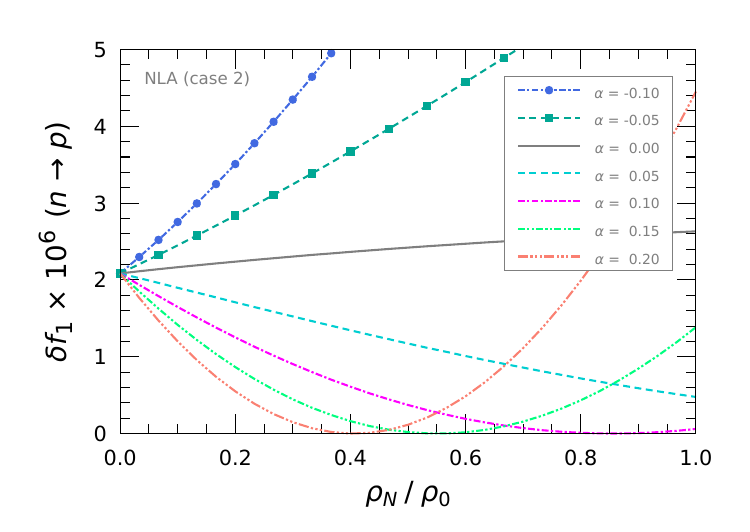} \\ %
  \includegraphics[width=8.0cm,keepaspectratio,clip,page=1]{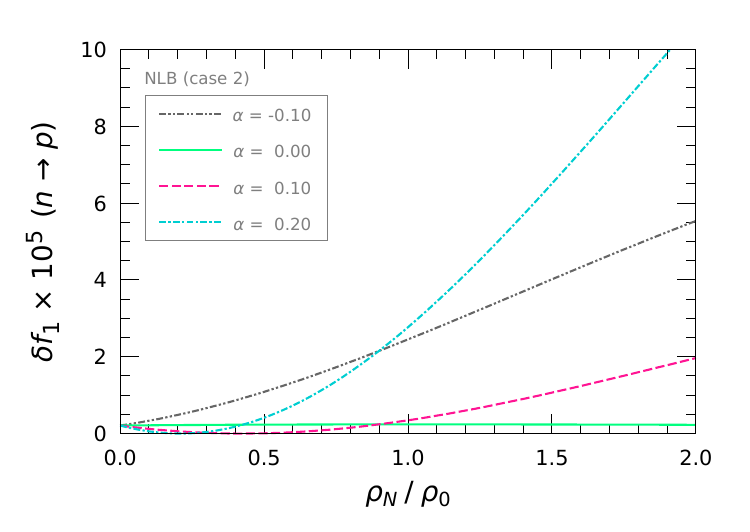}%
  \includegraphics[width=8.0cm,keepaspectratio,clip,page=1]{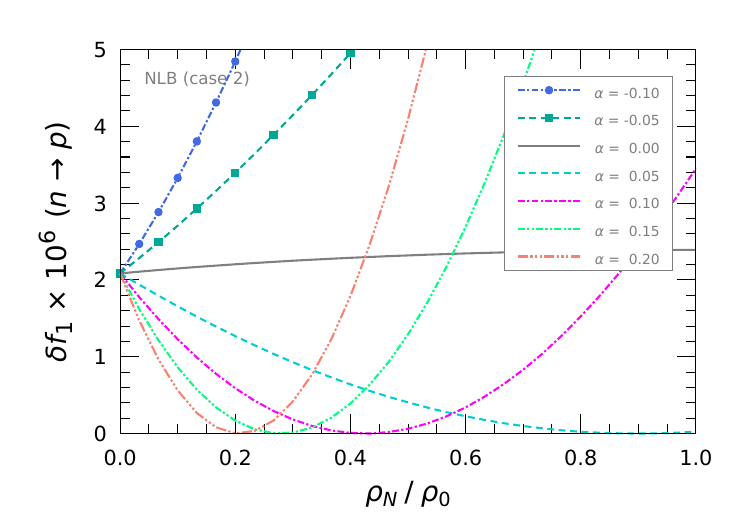} \\ %
  \includegraphics[width=8.0cm,keepaspectratio,clip,page=1]{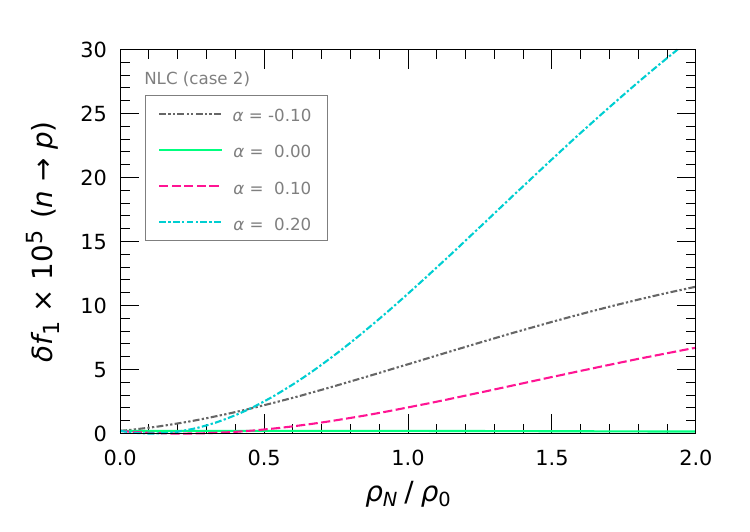}%
    \includegraphics[width=8.0cm,keepaspectratio,clip,page=1]{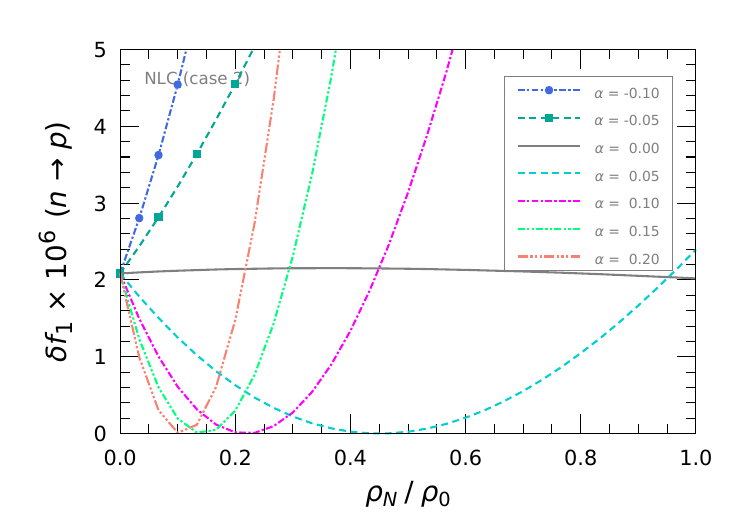}%
  \caption{\label{fig:FF-vector}
   Deviation, $\delta f_1$, for the neutron $\beta$ decay in the NLA, NLB or NLC model.  We show the results in case 2.  In the left panels, the nuclear density varies up to $\rho_N/\rho_0 = 2.0$.  In the right panels, $\delta f_1$ at low density is magnified. 
   }
\end{figure*}

In Fig.~\ref{fig:FF-vector}, we depict the density dependence of $\delta f_1$ for the neutron $\beta$ decay in the NLA, NLB or NLC model.  In proton-rich matter ($\alpha < 0$), $\delta f_1$ is simply enhanced with increasing the nuclear density, because $\Delta_{du}^\ast$ increases as $\rho_N$ grows up (see Fig.~\ref{fig:Mass-diff}).  In case of $\alpha = -0.1$, $\delta f_1$ reaches ${\cal O}(10^{-5})$ at $\rho_0$.  On the other hand, in neutron-rich matter ($\alpha > 0$), $\delta f_1$ decreases for low density and reaches zero at some density.  For example, in the NLB model with $\alpha = 0.2$, $\delta f_1$ vanishes at $\rho_N/\rho_0 \simeq 0.2$, at which $\Delta_{du}^\ast$ also vanishes  (see Fig.~\ref{fig:Mass-diff}) and isospin symmetry is restored.  Beyond such density, $\delta f_1$ begins to increase rapidly. 
We can see that, in the NLC model with $\alpha = 0.2$, $\delta f_1$ can reach ${\cal O}(10^{-4})$ at $\rho_0$, which is about $100$ times larger than the value in vacuum and is the same order as the current uncertainty in the $\beta$ decay experiments.  Therefore, although, at very low density, the effect of isovector scalar field on the weak vector coupling constant is small in neutron-rich nuclei, around $\rho_0$ it plays a significant role in the neutron $\beta$ decay in both neutron-rich and proton-rich nuclei.  

Next, we consider the hyperon decay.  Here we assume that a hyperon is simply immersed in asymmetric nuclear matter as an impurity, and that it does not affect the properties of nuclear matter~\cite{tsushima98the}.  In symmetric nuclear matter, we find that the effective mass of $\Lambda$ ($\Xi^-$) at $\rho_0$ is $1003 \, (1265)$ MeV, which shows the mass reduction of about $10 \%$ ($4 \%$) from the mass in vacuum~\cite{saito97var}, and that the mean-square charge radius of $\Lambda$ ($\Xi^-$) at $\rho_0$ is $0.035 \, (-0.64)$ fm$^2$, which is swelled in matter. This is similar to the case of proton in matter (see  section~\ref{sec:qmcmodel}). 

\begin{figure*}[h!]
  \includegraphics[width=8.0cm,keepaspectratio,clip,page=1]{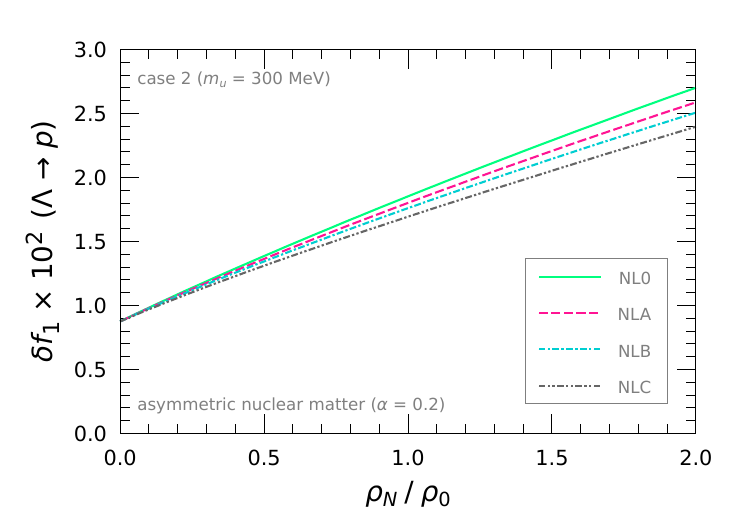}%
  \includegraphics[width=8.0cm,keepaspectratio,clip,page=1]{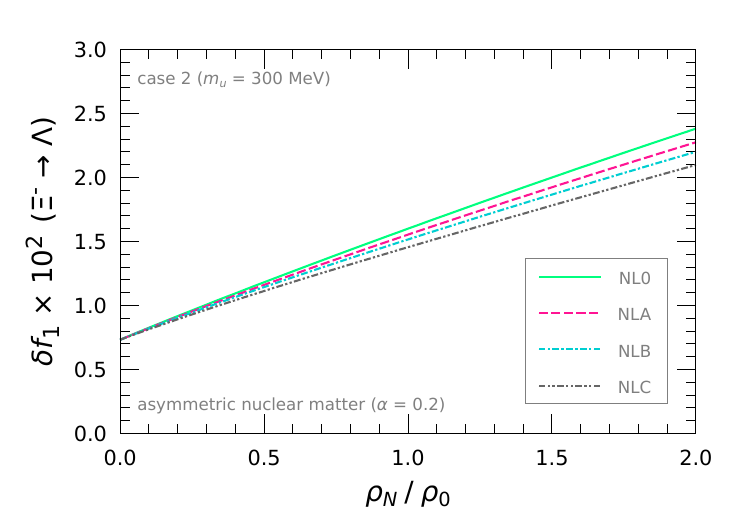} \\%
  \includegraphics[width=8.0cm,keepaspectratio,clip,page=1]{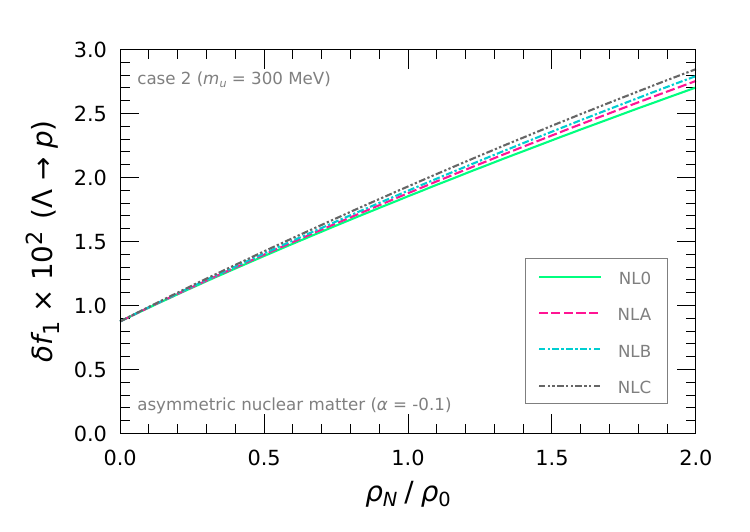}%
  \includegraphics[width=8.0cm,keepaspectratio,clip,page=1]{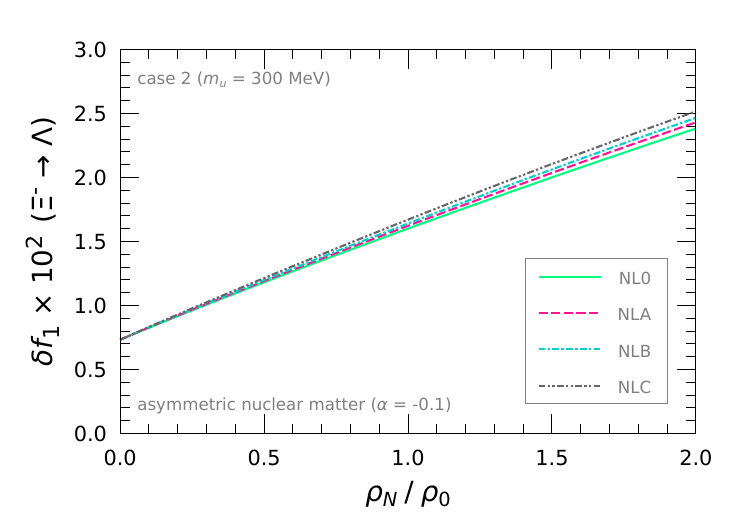}%
  \caption{\label{fig:FF-vector-Hyp}
   Deviation, $\delta f_1$, for the hyperon $\beta$ decay in the NL0, NLA, NLB or NLC model.  We show the results in case 2 only.  The left (right) two panels are for the $\Lambda$ ($\Xi^-$) decay. The top (bottom) two panels are for $\alpha = 0.2 \, (-0.1)$.
   }
\end{figure*}

The density dependence of $\delta f_1$ for the hyperon $\beta$ decay in matter is displayed in Fig.~\ref{fig:FF-vector-Hyp}. 
In the hyperon decay, the $s$ quark is changed into the $u$ quark, and the $u$-$s$ quark mass difference in matter is given by 
\begin{equation}
 \Delta_{su}^\ast = m_s^\ast - m_u^\ast = \Delta_{su} + \frac{g_\sigma}{3} {\bar \sigma}  + g_\delta {\bar \delta} , 
  \label{eq:sumassdiff}
\end{equation}
with $\Delta_{su} = 200$ MeV the mass difference in vacuum.  Therefore, the $\sigma$ field enhances the mass difference, while the $\delta$ field decreases (increases) $\Delta_{su}^\ast$ in neutron-rich (proton-rich) matter.  However, because the $\delta$ field may be weaker than the $\sigma$ field (see Fig.~\ref{fig:mean-fields}), $\Delta_{su}^\ast$ is eventually enhanced in asymmetric nuclear matter, which means that flavor SU(3) symmetry is {\em more} broken in matter than in vacuum.  Thus, $\delta f_1$ increases gradually as the density grows up.  We find that  $\delta f_1$ for the $\Lambda$ decay is larger than that for $\Xi^-$ decay, which is due to the cm correction $\rho_V$ as discussed in section~\ref{sec:weakff}. 
Because, in neutron-rich matter, the absolute value of the $\delta$ field in the NLC model is larger than that in the NLA model, SU(3) breaking in the NLC model is smaller  than that in the NLA model.  Thus, $\delta f_1$ in the NLC model is smaller than in the NLA model.  In contrast, in proton-rich matter, the situation is opposite, that is, $\delta f_1$ in the NLC model is larger than in the NLA model.  In the hyperon $\beta$ decay, the effect of the $\delta$ field is eventually not so remarkable, because $\Delta_{su}^\ast$ is almost determined by the isoscalar $\sigma$ field.

\subsection{Axial-vector coupling constant in matter} \label{subsec:axvectormatter}

\begin{figure*}[h!]
  \includegraphics[width=8.0cm,keepaspectratio,clip,page=1]{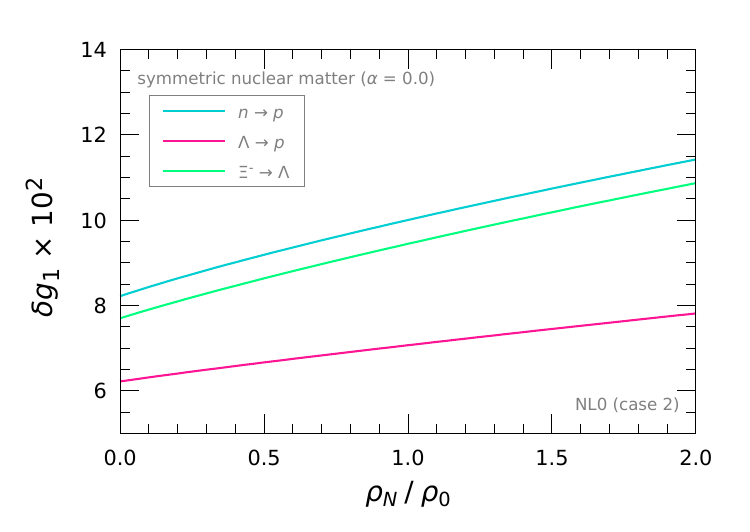}%
  \includegraphics[width=8.0cm,keepaspectratio,clip,page=1]{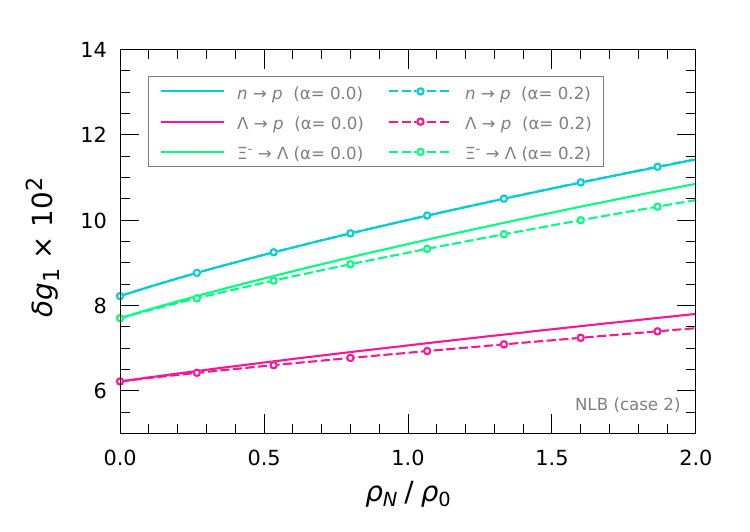}%
  \caption{\label{fig:FF-axial}
   Deviation, $\delta g_1$, in the axial-vector coupling constant as a function of $\rho_N/\rho_0$.  We show the results in case 2.  In the left panel, $\delta g_1$ in the NL0 model is displayed, in which the $\delta$ field is not included. 
   Because the dependence of $\delta g_1$ on $\alpha$ is quite small in this model, we show the result with $\alpha = 0$ only.  In the right panel, $\delta g_1$ in the NLB model is presented. In the neutron $\beta$ decay, the results with $\alpha = 0$ and $0.2$ are very close to each other. 
   }
\end{figure*}

We show the results of the axial-vector coupling constants in Fig.~\ref{fig:FF-axial}. Here we again notice that, because $\delta g_1$ is mainly generated by the relativistic effect of the quark wave function, the effective quark mass itself plays an essential role in $g_1$.  From the figure we see that the deviation, $\delta g_1$, is of the order of $10^{-2} \sim 10^{-1}$ and grows up gradually with increasing the density.  As in vacuum, we again find $\delta g_1(\Lambda \to p) < \delta g_1(\Xi^- \to \Lambda) < \delta g_1(n \to p)$.  

In symmetric nuclear matter, because the $\delta$ field is very small (see Fig.~\ref{fig:mean-fields}), the $u$ and $d$ quark masses are close to each other, namely $m_u^\ast \approx m_d^\ast$, and they are almost determined by the $\sigma$ field only.  On the other hand, in neutron-rich matter, due to the $\delta$ field, $m_d^\ast$ is lighter than in symmetric nuclear matter, while $m_u^\ast$ is heavier than in symmetric nuclear matter (see Fig.~\ref{fig:Quark-mass}). In proton-rich matter, the situation is opposite.  In spite of such modification on the quark mass, because $\delta g_1$ for the neutron decay is partly given by the overlap of the lower components of the $u$ and $d$ quark wave functions where one is enhanced and the other is reduced by the $\delta$ field, $\delta g_1$ in asymmetric nuclear matter is eventually not changed much from the value in symmetric nuclear matter, that is, the most effect of the $\delta$ field on $\delta g_1$ is canceled.  Thus, $\delta g_1$ for the neutron decay rarely depend on $\alpha$ (see the right panel of Fig.~\ref{fig:FF-axial}). 

On the other hand, in the hyperon $\beta$ decay, $\delta g_1$ in neutron-rich matter is smaller than that in symmetric nuclear matter.  This is because the $s$ quark mass does not change and $m_u^\ast$ is heavier than in symmetric nuclear matter, namely the lower component of the $u$ quark wave function is reduced. 

In any case, the effect of the $\delta$ field is relatively small and the isoscalar $\sigma$ field plays a vital role in the axial-vector coupling constant in matter (see also Ref.~\cite{saito95var}).  

\newpage

\clearpage

\section{Summary and conclusion} \label{sec:sumconc}

Using the quark-meson coupling (QMC) model~\cite{guichon88a, saito94a, saito07nucleon, guichon18qmc, krein18nucl}, we have investigated the effect of isoscalar and isovector,  scalar-mean fields on the breaking of SU(3) symmetry in asymmetric nuclear matter, and have calculated the weak vector and axial-vector coupling constants for semileptonic baryon decays in matter.  To avoid the radius problem~\cite{donoghue75phen}, which occurs in the bag model, we have used a relativistic confinement potential of the scalar-vector harmonic oscillator type in the QMC model.  In the present calculation, we have especially paid careful attention to the center of mass corrections to the baryon mass and the quark currents. 

First, we have considered the weak coupling constants in vacuum.  We introduce the constituent quark masses for the $u$, $d$ and $s$ quarks ($m_u < m_d \ll m_s$), which, of course, break SU(3) symmetry explicitly.  Then, those masses and other parameters appeared in the relativistic quark model are determined so as to reproduce the masses of proton, neutron, two hyperons ($\Lambda$ and $\Xi^-$) and the proton charge radius in vacuum.  In the present calculation, the $u$-$d$ quark mass difference, $\Delta_{du} \, (= m_d - m_u)$, is a few MeV, while the $u$-$s$ quark mass difference, $\Delta_{su}$, is $200$ MeV.  
Then, we have found that the deviation of the vector coupling constant from the value in exact SU(3) symmetry, $\delta f_1$, is given by $\delta f_1(n \to p) \sim {\cal O}(10^{-6}) \ll \delta f_1(\Xi^- \to \Lambda) < \delta f_1(\Lambda \to p) \sim {\cal O}(10^{-2})$, where $\delta f_1(n \to p) \sim {\cal O}(10^{-6})$ is the same order as in the original estimate~\cite{behrends60eff}. In contrast, the deviation for the axial-vector coupling constant, $\delta g_1$, is estimated as $\delta g_1(\Lambda \to p) \sim {\cal O}(10^{-2}) < \delta g_1(\Xi^- \to \Lambda) < \delta g_1(n \to p) \sim {\cal O}(10^{-1})$.  It is important to notice that the origin of $\delta f_1$ is the breaking of SU(3) symmetry, namely $\Delta_{du}$ or $\Delta_{su}$, while the main cause of $\delta g_1$ is the relativistic effect due to the lower component of the quark wave function.  
Furthermore, the cm corrections to the quark currents, $\rho_V$ and $\rho_A$, tend to reduce $\delta f_1$ and $\delta g_1$, respectively. 

Next, we have considered the weak coupling constants in asymmetric nuclear matter.  In the present calculation, we have used the QMC model, where the isovector scalar ($\delta$) meson as well as the isoscalar scalar ($\sigma$) and the vector ($\omega$ and $\rho$) mesons are introduced to describe the properties of nuclear matter.  Because, following the OZI rule, the $\sigma$ field couples to both the $u$ and $d$ quarks equally but does not couple to the $s$ quark, the $u$ and $d$ quark masses are reduced but the $s$ quark mass does not change in matter. Therefore, the $\sigma$ field keeps isospin (SU(2)) symmetry, but breaks SU(3) symmetry. In contrast, the $\delta$ field couples to the $u$ and $d$ quarks {\em oppositely}, that is, it increases (decreases) the $u$ ($d$) quark mass in neutron-rich matter, while it increases (decreases) the $d$ ($u$) quark mass in proton-rich matter.  It does not change the $s$ quark mass. Thus, the $\delta$ field breaks both isospin symmetry and SU(3) symmetry in matter.  

Now we have the breaking of SU(3) symmetry in matter as follows (see Eqs.~(\ref{eq:dumassdiff}) and (\ref{eq:sumassdiff})): 
\begin{align}
 \Delta_{du}^\ast &= m_d^\ast - m_u^\ast = \Delta_{du} + 2 g_\delta {\bar \delta}  ,  \nonumber \\
 \Delta_{su}^\ast &= m_s^\ast - m_u^\ast = \Delta_{su} + \frac{g_\sigma}{3} {\bar \sigma}  + g_\delta {\bar \delta} , \nonumber 
\end{align}
where $\Delta_{du}$ and $\Delta_{du}$ are the original breaking in vacuum, and the meson-field terms are the additional ones in matter.  It should be noticed that the mean field ${\bar \delta}$ is negative (positive) in neutron-rich (proton-rich) matter.  Because the amount of $\Delta_{du}$ is at most a few MeV and ${\bar \delta} < 0$ in neutron-rich matter, $\Delta_{du}^\ast$ vanishes at a certain low density (for example, $\rho_N/\rho_0 \sim 0.1 - 0.4$ in the NLA, NLB or NLC model with $\alpha = 0.2$), at which {\em isospin symmetry is thus restored and $\delta f_1$ in the neutron $\beta$ decay vanishes as well}.  Beyond such density, the absolute value of ${\bar \delta}$ grows up rapidly, and thus isospin symmetry is again broken. In the NLB or NLC model with $\alpha = 0.2$, we find that $\delta f_1(n \to p) \sim {\cal O}(10^{-5})$ or ${\cal O}(10^{-4})$ at $\rho_0$, respectively, which is about $10 -100$ times larger than the value in vacuum.  This fact implies that the effect of the isovector scalar field is quite vital in the analyses of the experiments of neutron $\beta$ decay in neutron-rich nuclei.  In contrast, in proton-rich matter, $\Delta_{du}^\ast$ increases monotonously as the density grows up because the $\delta$ field is positive, and thus the $\delta$ field also contributes to $\delta f_1$ considerably. 

In the hyperon decay in matter, the isoscalar $\sigma$ field simply enhances $\Delta_{su}^\ast$, while the isovector $\delta$ field reduces (enhances) $\Delta_{su}^\ast$ in neutron-rich (proton-rich) matter.  However, the effect of the $\delta$ field on $\delta f_1$ is eventually not so remarkable.  We find that $\delta f_1 (\Xi \to \Lambda) \sim {\cal O}(10^{-2}) < \delta f_1 (\Lambda \to p)$.  

Concerning the deviation for the axial-vector coupling constant, $\delta g_1$, it is rather sensitive to what kind of the quark model we choose.  Because we have adopted the relativistic constituent quark model in the present calculation, $\delta g_1$ is relatively small, compared with the value in the bag model~\cite{chodos74bag, cbm81, tonyadv13, saito95var}.  Here, we have investigated how the $\sigma$ and $\delta$ fields contribute to $\delta g_1$ qualitatively.  In matter, $\delta g_1$ is affected by the $\sigma$ field, but the effect of the $\delta$ field is not so large.  We have finally obtained that $\delta g_1(\Lambda \to p) \sim {\cal O}(10^{-2}) < \delta g_1(\Xi^- \to \Lambda) < \delta g_1(n \to p) \sim {\cal O}(10^{-1})$, which is similar to the result in vacuum.  

Now we add some caveats concerning the present results: (1) To focus our attention on the role of the isoscalar and isovector scalar fields in the weak vector and axial-vector coupling constants in matter, we have ignored the explicit degrees of freedom of pionic cloud and gluons in a baryon.  However, for any truely quantitative study of SU(3) symmetry breaking, such degrees of freedom should be considered simultaneously.  We can expect that such effects provide additional SU(3) breaking of the order of ${\cal O}(10^{-5}) \sim {\cal O}(10^{-4})$~\cite{kaiser01iso, crawford22charge}.  (2) We have assumed that the strength of the confinement potential, $c$, is constant.  This assumption should be also improved in the future consideration. (3) To compare the experiments of semileptonic baryon decays in nuclei, it is vital to compute the weak vector and axial-vector coupling constants in {\em finite nuclei}. 

We summarize by saying that the isovector scalar field considerably contributes to the breaking of SU(3) symmetry in nuclear matter. In particular, the size of the defect of the vector normalization for neutron $\beta$ decay in neutron-rich or proton-rich matter can reach the order of about $10^{-4}$ around the nuclear saturation density. This is close to the order of the current uncertainty in the measurements, and thus it would be necessary to require that the effect of the isovector scalar field be included in the future analyses. Theoretically, it is very interesting that, in neutron-rich nuclei, there exists a certain low density at which isospin symmetry is {\em restored}, namely $\Delta_{du}^\ast = 0$.  Furthermore, because the isoscalar scalar field reduces the $u$ and $d$ quark masses, it plays a significant role in the axial-vector coupling constant in matter.

\section*{Acknowledgments}

This work was supported by the National Research Foundation of Korea (Grant Nos. RS-2023-00242196, NRF-2021R1A6A1A03043957 and NRF-2020R1A2C3006177).
\newpage

\appendix

\section{\label{app:cm} The center of mass corrections}

\subsection{\label{app:mass} The cm correction to the baryon mass}

Following the method described in Ref.~\cite{guichon96role}, we estimate the cm correction to the baryon mass.  

Assuming that the Hamiltonian density ${\cal H}_B$ for a composite baryon can be separated  into the internal and the center of mass parts as ${\cal H}_B = {\cal H}_B^{in} + {\cal H}_B^{cm}$, we first define  
\begin{align}
  {\vec R} 
  & = \frac{1}{3} \sum_i {\vec r}_i , \ \ \ {\vec P} = \frac{1}{3} \sum_i {\vec p}_i , \\
  {\vec \rho}_i  
  & = {\vec r}_i - {\vec R}  ,  \ \ \  {\vec \pi}_i = {\vec p}_i - {\vec P} , \ \ \  \sum_i {\vec \rho}_i = {\vec 0} , \ \ \   \sum_i {\vec \pi}_i = {\vec 0} ,  
\end{align}
where ${\vec r}_i$ and ${\vec p}_i$ are respectively the position and momentum of quark $i$.  The Hamiltonian density ${\cal H}_B$ is then given by 
\begin{align}
  {\cal H}_B &= \sum_i \gamma_0(i) \left[ {\vec \gamma}(i) \cdot {\vec p}_i + m_i + U({\vec r}_i) \right] ,   \label{eq:hamd}  \\
  &= \sum_i \gamma_0(i) \left[ {\vec \gamma}(i) \cdot {\vec \pi}_i + m_i + U({\vec \rho}_i) \right]   
   + \sum_i \gamma_0(i) \left[ {\vec \gamma}(i) \cdot {\vec P}_i + U({\vec \rho}_i + {\vec R}) - U({\vec \rho}_i) \right] ,    \\
  & \equiv {\cal H}_B^{in} + {\cal H}_B^{cm}
\end{align}
with $U({\vec r}_i) = \frac{c}{2} \left( 1+\gamma_0(i) \right) {\vec r}_i^{\, 2}$. 
Using the eigenstate of ${\cal H}_B$, $| B \rangle$,  the cm correction to the baryon energy is calculated by 
\begin{align}
E_B^{cm} &= \langle B| {\cal H}_B^{cm} | B \rangle = 
\langle B| \sum_i \gamma_0(i) \left[ {\vec \gamma}(i) \cdot {\vec P} + U({\vec \rho}_i + {\vec R}) - U({\vec \rho}_i) \right] | B \rangle ,    \\
  & \equiv \langle B| {\cal H}_B^{cm(1)} | B \rangle + \langle B| {\cal H}_B^{cm(2)} | B \rangle = E_B^{cm(1)} + E_B^{cm(2)} 
\end{align}
with 
\begin{align}
{\cal H}_B^{cm(1)} &=  {\vec P} \cdot \sum_i \gamma_0(i) {\vec \gamma}(i)  = \frac{1}{3} \sum_j {\vec p}_j \cdot \sum_i \gamma_0(i) {\vec \gamma}(i) , \\
{\cal H}_B^{cm(2)} &=  \frac{c}{2} \sum_i (1+\gamma_0(i)) ( {\vec r}_i^{\, 2} -  {\vec \rho}_i^{\, 2} ) = \frac{c}{2} \sum_i (1+\gamma_0(i)) ( 2 {\vec r}_i \cdot {\vec R} -  {\vec R}^{\, 2} ) . 
\end{align}

Let $| i \rangle$ be the lowest, single-particle solution with energy $\epsilon_i$, namely 
\begin{equation}
\left[ {\vec \alpha} \cdot {\vec p}_i + \gamma_0(i) m_i + \frac{c}{2} \left( 1+ \gamma_0(i) \right) {\vec r}_i^{\, 2}  \right] |i \rangle = \epsilon_i  |i \rangle  . 
\end{equation}
When all the quarks in baryon are in the ground state, we find  
\begin{align}
E_B^{cm(1)} &=  \frac{1}{3} \sum_{i, j} \langle B| {\vec p}_j \cdot \gamma_0(i) {\vec \gamma}(i) | B \rangle = \frac{1}{3} \sum_i \langle B| {\vec p}_i \cdot {\vec \alpha}(i) | B \rangle  \nonumber  \\
&= \frac{1}{3} \sum_i \langle i| \epsilon_i - \gamma_0(i) m_i - \frac{c}{2} \left( 1+ \gamma_0(i) \right) {\vec r}_i^{\, 2}  | i \rangle = \frac{1}{3} \sum_i \frac{6}{a_i^2 ( 3\epsilon_i + m_i)} . 
\end{align}
Similarly, we can calculate $E_B^{cm(2)}$ as follows.  Defining 
\begin{equation}
E_B^{cm(2)} = c \sum_i \langle B| (1+\gamma_0(i)) {\vec r}_i \cdot {\vec R} | B \rangle  
- \frac{c}{2} \sum_i \langle B| (1+\gamma_0(i)) {\vec R}^2 | B \rangle  \equiv 
E_B^{cm(2a)}  - E_B^{cm(2b)} , 
\end{equation}
$E_B^{cm(2a)}$ reads 
\begin{align}
E_B^{cm(2a)} &=  \frac{c}{3} \sum_{i, j} \langle B| (1+\gamma_0(i)) {\vec r}_i \cdot {\vec r}_j | B \rangle  = \frac{c}{3} \sum_{i} \langle B| (1+\gamma_0(i)) {\vec r}_i^{\, 2} | B \rangle  ,  \nonumber \\
&= \frac{c}{3} \sum_i \langle i| \left( 1+ \gamma_0(i) \right) {\vec r}_i^{\, 2}  | i \rangle 
= \frac{1}{3} \sum_i \frac{6}{a_i^2 ( 3\epsilon_i + m_i)} , 
\end{align}
and $E_B^{cm(2b)}$ reads 
\begin{align}
E_B^{cm(2b)} &=  \frac{c}{2} \sum_i \langle B| (1+\gamma_0(i)) {\vec R}^2 | B \rangle 
= \frac{c}{18} \sum_{i, j, k} \langle B| (1+\gamma_0(i))  {\vec r}_j \cdot {\vec r}_k | B \rangle , \nonumber  \\
&= \frac{c}{18} \sum_i \langle i| \left( 1+ \gamma_0(i) \right) {\vec r}_i^{\, 2}  | i \rangle 
+ \frac{c}{18} \sum_i \sum_{j \neq i} \langle i| \left( 1+ \gamma_0(i) \right) | i \rangle \langle j| {\vec r}_j^{\, 2}  | j \rangle ,  \nonumber \\
&= \frac{1}{6} E_B^{cm(2a)} + \frac{c}{3} \sum_i \sum_{j \neq i} \frac{\epsilon_i + m_i}{3\epsilon_i + m_i} \cdot \frac{11\epsilon_{j}+m_{j}}{\left(3\epsilon_{j}+m_{j}\right)\left(\epsilon_{j}^2-m_{j}^2\right)}. 
\end{align}

Thus, the total cm correction is 
\begin{equation}
  E^{cm}_{B}
  =\frac{11}{3}\sum_{i}A_{i}
  +\frac{1}{9}\sum_{i}\sum_{j} \left( \delta_{ij} - 1 \right) C_{i}D_{j},
  \label{eq:Ecm1}
\end{equation}
with
\begin{equation}
  A_{i}
  =\frac{1}{a_{i}^{2}(3\epsilon_{i}+m_{i})}, \ \ \ 
  C_{i}
  =\frac{\epsilon_{i}^2-m_{i}^2}{a_{i}^{2}(3\epsilon_{i}+m_{i})}=\frac{3}{a_{i}^{2}}A_{i}, \ \ \ 
  D_{i}
  =\frac{11\epsilon_{i}+m_{i}}{\left(3\epsilon_{i}+m_{i}\right)\left(\epsilon_{i}^2-m_{i}^2\right)}.  
  \label{eq:acd1}
\end{equation}
When all the quarks in baryon have the same mass, the correction is 
\begin{equation}
  E^{cm}_{B}
  = \frac{77\epsilon + 31 m}{3a^2(3\epsilon+m)^2}  , 
  \label{eq:Ecmsym}
\end{equation}
which is exactly the same as in Refs.~\cite{barik13n, zhu23eos}.

\subsection{\label{app:radius} The mean-square charge radius of baryon}

As in Appendix~\ref{app:mass}, we calculate the mean-square charge radius of baryon including the cm correction.  It is defined by
\begin{equation}
\langle r^2 \rangle_B = \sum_i e_i \langle B| ({\vec r} - {\vec R})^2 |B \rangle = \sum_i e_i \langle B| {\vec r}^{\, 2} - 2 {\vec r} \cdot {\vec R} + {\vec R}^{\, 2} |B \rangle ,
\end{equation}
with $e_i$ the fractional charge of quark $i$.  Using $\langle i| {\vec r}_i^{\, 2}  | i \rangle = 3D_i/2$ (see Eq.~(\ref{eq:acd1})), the radius is then calculated as
\begin{equation}
\langle r^2 \rangle_B = \frac{1}{2} \sum_i e_i D_i + \frac{Q_B}{6} \sum_i D_i ,  \label{eq:chargeradius}
\end{equation}
with $Q_B$ the total charge of baryon.  
For instance, we respectively find the charge radii of proton, neutron, $\Lambda$ and $\Xi^-$ as
\begin{align}
  \langle r^2 \rangle_{p} &=D_{u} , \ \ \ 
  \langle r^2 \rangle_{n} =\frac{1}{3} \left( D_{u} - D_d \right)  ,   \nonumber \\
  \langle r^2 \rangle_{\Lambda} &=\frac{1}{6} \left( 2D_{u} - D_d - D_s \right)  , \ \ \ 
  \langle r^2 \rangle_{\Xi^-} =-\frac{1}{3} \left( D_{d} + 2D_s \right) . 
  \label{eq:radiiinapp}
\end{align}

\subsection{\label{app:current} The cm corrections to the vector and axial-vector currents}

Following Wong~\cite{wong81cm} and Bartelski et al.~\cite{bartelski84cm}, the better approximation to estimate the cm correction to the quark current may be to consider the quark-model wave function (at the origin), $| B, s \rangle_{\vec 0}$, as a wave packet of the physical particle 
\begin{equation}
  | B, s \rangle_{\vec 0} =  \int d{\vec p} \, \phi_B(|{\vec p}|) \, |{\vec p}, s \rangle . 
  \label{eq:wavepacket}
\end{equation}
The wave packet on the right-hand side of Eq.~(\ref{eq:wavepacket}) is assumed to have the same parity and spin $s$ as the static, quark-model state on the left.  The weighting function $\phi_B$ is a function of $|{\vec p}\,|$ only.   Here, $|{\vec p}, s \rangle$ is the eigenstate of momentum defined by 
\begin{equation}
  \braket{ {\vec r} \, | {\vec p}, s} = \sqrt{\frac{E_B + M_B}{2E_B}} 
    \begin{pmatrix}
1 \\
\frac{{\vec \sigma} \cdot {\vec p}}{E_B + M_B}  
\end{pmatrix}
  e^{i {\vec p}\cdot {\vec r}} \chi_s  \equiv u_B({\vec p}\,) e^{i {\vec p}\cdot {\vec r}} \chi_s  ,  
  \label{eq:planewave}
\end{equation}
where $M_B$ is the mass of baryon $B$ and $E_B = \sqrt{M_B^2 + {\vec p\,}^2}$.  This plane wave state is normalized as $\braket{ {\vec p\,}^\prime, s^\prime | {\vec p}, s} = (2\pi)^3 \delta ({\vec p} - {\vec p}^\prime) \delta_{s, s^\prime}$. 

Applying the translation operator $e^{i{\vec P}\cdot {\vec b}/2}$ (with ${\vec P}$ the  momentum operator), we find 
\begin{equation}
  | B, s \rangle_{{\vec b}/2} =  \int d{\vec p} \, \phi_B(p) \, e^{i{\vec p}\cdot {\vec b}/2} \, |{\vec p}, s \rangle . 
  \label{eq:wavepacket2}
\end{equation}
 We then calculate the Hill-Wheeler overlap as 
\begin{align}
  _{-{\vec b}/2}\braket{B, s^\prime | B, s }_{{\vec b}/2} &=  \int d{\vec p\,}^\prime \int d{\vec p} \, \braket{ {\vec p\,}^\prime, s^\prime | \phi_B^\ast(p^\prime) \, e^{i{\vec p\,}^\prime \cdot {\vec b}/2} \, e^{i{\vec p} \cdot {\vec b}/2} \phi_B(p) |{\vec p}, s} ,  \nonumber 
  \\
  &= (2\pi)^3 \delta_{s, s^\prime} \int d{\vec p} \, |\phi_B(p)|^2 \, e^{i{\vec p}\cdot {\vec b}} . 
  \label{eq:HWoverlap}
\end{align}
Taking the inverse Fourier transformation of the overlap, we obtain 
\begin{equation}
 I_B(p) 
 = \frac{1}{(2\pi)^3} \int d{\vec b} \ 
  { _{-{\vec b}/2}\braket{B, s^\prime | B, s }_{{\vec b}/2} } \ e^{-i{\vec p}\cdot {\vec b}} = (2\pi)^3 |\phi_B(p)|^2 , 
  \label{eq:HWI}
\end{equation}
which is normalized as 
\begin{equation}
 \int d{\vec p} \ I_B(p) = 1 , 
  \label{eq:HWI1}
\end{equation}
and we find
\begin{equation}
 \phi_B(p) 
 = \frac{1}{(2\pi)^{3/2}} \sqrt{I_B(p)} . 
  \label{eq:phi}
\end{equation}

Using the quark wave function given by Eq.~(\ref{eq:sol}), the overlap function can be calculated by
\begin{equation}
  J_i(b) =  \int d{\vec r} \, \psi_i^\dagger\left({\vec r}+ {\vec b}/{2} \right) \psi_i\left({\vec r}- {\vec b}/{2}\right) 
   = \left( 1- \frac{b^2}{a_i^2} d_i \right) e^{-b^2/4a_i^2} , 
  \label{eq:qoverlap}
\end{equation}
with $d_i = 1/(4\lambda_i^2a_i^2 + 6)$. 
Therefore, when the baryon consists of three quarks with flavor $i$, $j$ and $k$, we find 
\begin{align}
  I_B&(p) = \frac{1}{(2\pi)^3} \int d{\vec b} \, J_i(b)J_j(b)J_k(b) e^{-i{\vec p}\cdot {\vec b}}  \nonumber \\
    &= \frac{a_0^3}{(3\pi)^{3/2}} \left[ 1 - 2a_0^2 F_1 + \frac{20}{3} a_0^4 F_2 - \frac{280}{9} a_0^6 F_3  
  + \left(  \frac{4}{9} a_0^2 F_1 - \frac{80}{27} a_0^4 F_2 + \frac{560}{27} a_0^6 F_3 \right) (a_0p)^2  \right. \nonumber  \\ 
    &  \left.  + \left( \frac{16}{81} a_0^4 F_2 - \frac{224}{81} a_0^6 F_3 \right) (a_0p)^4 
    + \left( \frac{8}{27} \right)^2 a_0^6 F_3 (a_0p)^6  \right] e^{-a_0^2p^2/3}  ,       
  \label{eq:Iijk}
\end{align}
where 
\begin{align}
\frac{1}{a_0^2} & = \frac{1}{3} \left(  \frac{1}{a_i^2} + \frac{1}{a_j^2} + \frac{1}{a_k^2} \right) ,  \\ 
F_1 & = \frac{d_i}{a_i^2} + \frac{d_j}{a_j^2} + \frac{d_k}{a_k^2} ,  \ \ 
F_2  = \frac{d_id_j}{a_i^2a_j^2} + \frac{d_jd_k}{a_j^2a_k^2} + \frac{d_kd_i}{a_k^2a_i^2} ,  \ \ 
F_3  = \frac{d_id_jd_k}{a_i^2a_j^2a_k^2}  .  
  \label{eq:Iijk2}
\end{align}

Using the plane-wave expansion, Eq.~(\ref{eq:wavepacket}), the matrix elements of the decay $B_1 \to B_2 + l + {\bar \nu}_l$ read  
\begin{align}
 \int d{\vec r} \ \braket{B_2 | V^0(x) | B_1 } &=  \int d{\vec r} \int d{\vec p\,}^\prime \int d{\vec p} \, \phi_2^\ast(p^\prime) \, \phi_1(p) f_1(q^2) {\bar u}_2({\vec p}\,) \gamma^0 u_1({\vec p}\,) e^{i ({\vec p} - {\vec p\,}^\prime )\cdot {\vec r}} ,   \nonumber \\
& = f_1(0)  (2\pi)^3 \int d{\vec p} \, \phi_2^\ast(p) \phi_1(p) {\bar u}_2({\vec p}\,) \gamma^0 u_1({\vec p}\,) ,   \nonumber \\ 
& \equiv f_1 \times \rho_V(B_1 \to B_2)    ,       \label{eq:rhov}  \\
  \int d{\vec r} \ \braket{B_2 | A^3(x) | B_1 } &=  \int d{\vec r} \int d{\vec p\,}^\prime \int d{\vec p} \, \phi_2^\ast(p^\prime) \, \phi_1(p) g_1(q^2) {\bar u}_2({\vec p}\,) \gamma^3 \gamma_5 u_1({\vec p}\,) e^{i ({\vec p} - {\vec p\,}^\prime )\cdot {\vec r}} ,   \nonumber \\
& = g_1(0) (2\pi)^3 \int d{\vec p} \, \phi_2^\ast(p) \phi_1(p) {\bar u}_2({\vec p}\,) \gamma^3 \gamma_5 u_1({\vec p}\,)  ,  \nonumber \\ 
& \equiv g_1 \times \rho_A(B_1 \to B_2)  ,    \label{eq:rhoa}  
\end{align}
with 
\begin{align}
 \rho_V(B_1 \to B_2) & = (2\pi)^3 \int d{\vec p} \, \phi_2^\ast(p) \phi_1(p) {\bar u}_2({\vec p}\,) \gamma^0 u_1({\vec p}\,)    ,       \label{eq:defrhov}  \\
 \rho_A(B_1 \to B_2) & = (2\pi)^3 \int d{\vec p} \, \phi_2^\ast(p) \phi_1(p) {\bar u}_2({\vec p}\,) \gamma^3 \gamma_5 u_1({\vec p}\,)   ,    \label{eq:defrhoa}  
\end{align}
where $V^0(x)$ and $A^3(x)$ are respectively the vector and axial-vector currents (see Eqs.~(\ref{eq:weakv}) and (\ref{eq:weaka})), and $\phi_{1 (2)} = \phi_{B_1 (B_2)}$ etc.  Here, $f_1$ and $g_1$ are respectively the corrected, vector and axial-vector coupling constants, and the functions $\rho_V$ and $\rho_A$ are respectively the cm corrections to the uncorrected coupling constants, $f_1^{(0)}$ and $g_1^{(0)}$, calculated by the quark model (see Eqs.~(\ref{eq:gv0}) and (\ref{eq:ga0})).  
We thus find~\cite{donoghue87km} 
\begin{align}
 f_1 &= \frac{1}{\rho_V} \int d{\vec r} \ \braket{B_2 | V^0(x) | B_1 }  = \frac{f_1^{(0)}}{\rho_V} , 
 \label{eq:corgv}  \\
  g_1 &= \frac{1}{\rho_A} \int d{\vec r} \ \braket{B_2 | A^3(x) | B_1 } = \frac{g_1^{(0)}}{\rho_A} . 
 \label{eq:corga}  
\end{align}

Now we calculate the cm corrections explicitly.  Using Eq.~(\ref{eq:planewave}), we obtain 
\begin{align}
 \rho_V & = 16 \pi^4 \int_0^\infty dp \, p^2 \sqrt{\frac{E_1 + M_1}{E_1 E_2 (E_2 + M_2)}} \left( E_1 + E_2 - M_1 + M_2 \right) \phi_2^\ast(p) \phi_1(p)   ,       \label{eq:exrhov}  \\ 
  & \xrightarrow{{\rm if} \, M_1 = M_2}  \braket{1} = 1 , \nonumber \\ 
  \rho_A & = 16 \pi^4 \int_0^\infty dp \, p^2 \sqrt{\frac{(E_1 + M_1)(E_2 + M_2)}{E_1 E_2}} \left[ 1 - \frac{E_1 - M_1}{3(E_2 + M_2)} \right] \phi_2^\ast(p) \phi_1(p)  , \label{eq:exrhoa}  \\
   & \xrightarrow{{\rm if} \, M = M_1 = M_2}  \braket{1} - \frac{\braket{p^2}}{3M^2}  + \frac{\braket{p^4}}{4M^4} + \cdots  , \nonumber
\end{align}
where $M_{1 (2)}$ is the mass of baryon $B_{1 (2)}$, $E_{1 (2)} = \sqrt{p^2 + M_{1 (2)}^2}$, and $\braket{p^n}$ is the average of $p^n$ with respect to the weighting function $\phi_B \, (= \phi_1 = \phi_2)$ 
\begin{equation}
\braket{p^n} \equiv (2\pi)^3 \int d{\vec p} \, p^{n} |\phi_B(p)|^2 .  \label{eq:bracketp} 
\end{equation}

Here we show that $\rho_V$ follows the BSAG theorem.  As an example, we consider the neutron $\beta$ decay.  In this case, $\rho_V$ is rewritten as 
\begin{equation}
 \rho_V = \frac{1}{2} \int d{\vec p} \, \sqrt{\frac{E_n + M_n}{E_n E_p (E_p + M_p)}} \left( E_n + E_p - M_n + M_p \right) \sqrt{I_n(p) I_p(p)}   .       \label{eq:nrhov}  
\end{equation}
First, defining $\delta M = M_n - M_p$, $\delta E = E_n - E_p$, $M = M_p$ and $E = E_p$, we can expand 
\begin{align}
& \frac{1}{2} \sqrt{\frac{E_n + M_n}{E_n E_p (E_p + M_p)}} \left( E_n + E_p - M_n + M_p \right)  \nonumber \\ 
&= \frac{1}{2} \sqrt{\frac{E + M + \delta E + \delta M}{E (E + \delta E) (E + M)}} \left( 2E + \delta E - \delta M \right)  \nonumber \\
&= 
1 + \frac{E \delta E - M \delta M}{2E (E + M)} + {\cal O}(\delta M^2)  
= 1 +  {\cal O}(\delta M^2) = 1 +  {\cal O}(\Delta_{du}^2)  , 
   \label{eq:prefactor1}  
\end{align}
where we have used the relation, $E_n^2 - M_n^2 = E_p^2 - M_p^2$, and $\delta M$ is proportional to $\Delta_{du}$. 
Next, we consider the function $I_n(p)$.  From Eqs.~(\ref{eq:qoverlap}) and (\ref{eq:Iijk}), we find 
\begin{align}
  I_n(p) & = \frac{1}{(2\pi)^3} \int d{\vec b} \, J_u(b)J_d(b)J_d(b) e^{-i{\vec p}\cdot {\vec b}}  \nonumber \\
    &= \frac{1}{(2\pi)^3} \int d{\vec b} \, J_u(b)J_d(b)  \left[ \left( 1- \frac{d_d}{a_d^2} b^2 \right) e^{-b^2/4a_d^2} \right] e^{-i{\vec p}\cdot {\vec b}}  \nonumber \\    
    &= \frac{1}{(2\pi)^3} \int d{\vec b} \, J_u(b)J_d(b) \left[ J_u(b) - \left( h_1 b^2- h_2 b^4 \right) e^{-b^2/4a_u^2} + {\cal O}(\Delta_{du}^2) \right] e^{-i{\vec p}\cdot {\vec b}}  \nonumber \\    
    &= I_p(p) - \frac{1}{(2\pi)^3} \int d{\vec b} \, J_u(b)J_d(b) \left[ \left( h_1 b^2- h_2 b^4 \right) e^{-b^2/4a_u^2} \right] e^{-i{\vec p}\cdot {\vec b}} + {\cal O}(\Delta_{du}^2)  , 
  \label{eq:In}
\end{align}
where the functions $h_1$ and $h_2$ are of the order of $\Delta_{du}$.  We thus have 
\begin{align}
 \rho_V &= \int d{\vec p} \, \left[ 1 + {\cal O}(\Delta_{du}^2) \right] \sqrt{I_n(p) I_p(p)}  \nonumber \\
 &= \int d{\vec p} \, I_p(p)  - \frac{1}{2(2\pi)^3} \int d{\vec p} \int d{\vec b} \, J_u(b)J_d(b) \left[ \left( h_1 b^2- h_2 b^4 \right) e^{-b^2/4a_u^2} \right] e^{-i{\vec p}\cdot {\vec b}} + {\cal O}(\Delta_{du}^2)  \nonumber \\
 &= 1 + {\cal O}(\Delta_{du}^2)  ,       \label{eq:nrhov1}  
\end{align}
where the term of the order of $\Delta_{du}$ vanishes.  

\begin{figure*}[h!]
  \includegraphics[width=12cm,keepaspectratio,clip]{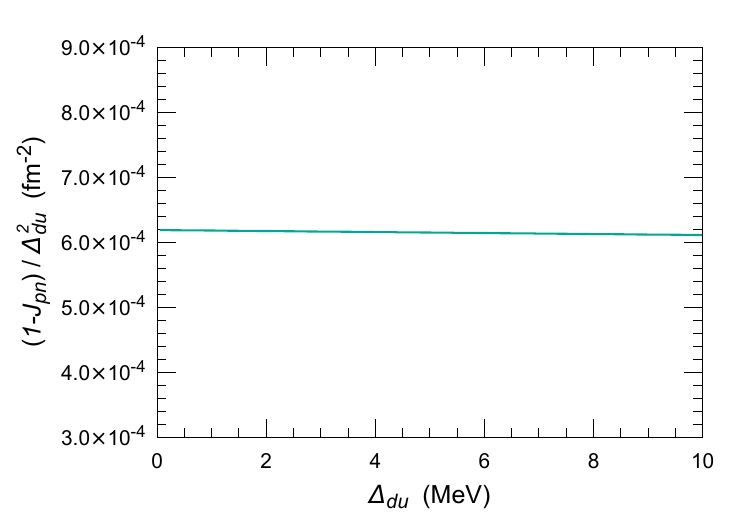}%
  \caption{\label{fig:Jpn}
    $J_{pn}$ is defined by Eq.~(\ref{eq:Jpn}). Here we take $m_u = 300$ MeV and vary the $d$ quark mass.
    }
\end{figure*}

Furthermore, defining $J_{pn}$ by 
\begin{equation}
 J_{pn} = \int d{\vec p} \, \sqrt{I_n(p) I_p(p)}   ,       \label{eq:Jpn}  
\end{equation}
we numerically check whether $J_{pn}$ follows the BSAG theorem.   In Fig.~\ref{fig:Jpn}, 
we depict $(1-J_{pn}) / \Delta_{du}^2$, which is constant for small $\Delta_{du}$. 

\begin{figure*}[h!]
  \includegraphics[width=12cm,keepaspectratio,clip]{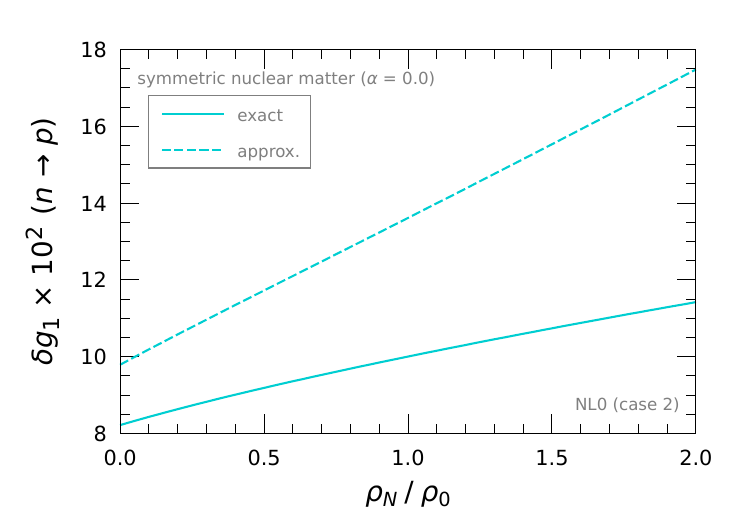}%
  \caption{\label{fig:approx}
    $\delta g_1$ for the neutron $\beta$ decay in symmetric nuclear matter.  We take the NL0 model with case 2. 
    }
\end{figure*}

Therefore, it is verified that the cm correction $\rho_V$ obeys the BSAG theorem.  
Because, as discussed in section~\ref{sec:weakff}., $f_1^{(0)}$ obeys the BSAG theorem, the corrected, vector coupling $f_1$ (Eq.~(\ref{eq:corgv})) obeys the BSAG theorem as well. 

Finally, the cm corrections are often approximated with the expansions in terms of $(p/M)^2$ as~\cite{bartelski84cm, donoghue87km} 
\begin{align}
 \rho_V \simeq \braket{1} & + \frac{\braket{p^2}}{4M_1M_2} \left[ 1- \frac{1}{2} \left( \frac{M_2}{M_1} + \frac{M_1}{M_2} \right) \right]  \nonumber \\
 & + \frac{\braket{p^4}}{128M_1^2M_2^2} \left[ 11\left( \frac{M_2^2}{M_1^2} + \frac{M_1^2}{M_2^2} \right)-12 \left( \frac{M_2}{M_1} + \frac{M_1}{M_2} \right) + 2 \right]    ,     
   \label{eq:aprhov}  \\ 
 \rho_A \simeq  \braket{1} & - \frac{\braket{p^2}}{3M_1M_2} \left[ \frac{1}{4} + \frac{3}{8} \left( \frac{M_2}{M_1} + \frac{M_1}{M_2} \right) \right]  \nonumber \\
 & + \frac{\braket{p^4}}{128M_1^2M_2^2} \left[ 11\left( \frac{M_2^2}{M_1^2} + \frac{M_1^2}{M_2^2} \right) + 4 \left( \frac{M_2}{M_1} + \frac{M_1}{M_2} \right) + 2 \right] . \label{eq:aprhoa}
\end{align}
As an example, in Fig.~\ref{fig:approx}, 
we show $\delta g_1$ in the neutron $\beta$ decay (see Eq.~(\ref{eq:corgva})) to compare the result calculated by the full form (Eq.~(\ref{eq:exrhoa})) with that by the approximate one (Eq.~(\ref{eq:aprhoa})).  As seen in the figure, the exact result is {\em not} well reproduced by the approximation.

\section{\label{app:asymprop} The binding energy per nucleon in asymmetric nuclear matter}

In general, the binding energy per nucleon, $E_b$, in asymmetric nuclear matter is defined as~\cite{chen09high} 
\begin{align}
  E_{b}(\rho_{N},\alpha)
  & = E(\rho_{N},\alpha) -\frac{1}{2}\left[(M_{n}+M_{p})+(M_{n}-M_{p})\alpha\right]
    \nonumber \\
  & = E_{0}(\rho_{N}) + E_{ISB}(\rho_{N})\alpha + E_{sym}(\rho_{N})\alpha^{2} + \mathcal{O}(\alpha^{3}),
    \label{eq:binding-energy1}
\end{align}
with $\alpha=(\rho_{n}-\rho_{p})/\rho_{N}$ and $E$ the total energy per nucleon (see Eq.~(\ref{eq:energydensity})).  Here, the isospin symmetry breaking (ISB) energy and the nuclear symmetry energy are given by
\begin{equation}
  E_{ISB}(\rho_{N}) = \left.\frac{\partial E(\rho_{N},\alpha)}{\partial\alpha}\right|_{\alpha=0},
  \quad
  E_{sym}(\rho_{N}) = \frac{1}{2}\left.\frac{\partial^{2}E(\rho_{N},\alpha)}{\partial\alpha^{2}}\right|_{\alpha=0}.
\end{equation}
Then, we expand $E_{0}(\rho_{N})$ and $E_{\rm sym}(\rho_{N})$ around the nuclear saturation density, $\rho_{0}$, as
\begin{align}
  &E_{0}(\rho_{N})
   = E_{0}(\rho_{0}) + \frac{K_{0}}{2}\chi^{2} + \frac{J_{0}}{6}\chi^{3} + \mathcal{O}(\chi^{4}),
    \label{eq:Binding-energy-sym} \\
  &E_{sym}(\rho_{N})
   = E_{sym}(\rho_{0}) + L \chi + \frac{K_{sym}}{2}\chi^{2} + \frac{J_{sym}}{3!}\chi^{3} + \mathcal{O}(\chi^{4}),
    \label{eq:Binding-energy-asym}
\end{align}
with $\chi=(\rho_{N}-\rho_{0})/3\rho_{0}$.  
Here, the incompressibility coefficient of symmetric nuclear matter, $K_{0}$, the slope and the curvature parameters of nuclear symmetry energy, $L$ and $K_{sym}$, and the third-order incompressibility coefficients, $J_{0}$ and $J_{sym}$, are respectively calculated as
\begin{align}
  K_{0} = 9 \rho_{N}^{2}
  & \left. \frac{d^{2}E_{0}(\rho_{N})}{d\rho_{N}^{2}} \right|_{\rho_{N}=\rho_{0}},
    \label{eq:definition-K0}\\
  L = 3 \rho_{N} \left. \frac{d E_{sym}(\rho_{N})}{d\rho_{N}} \right|_{\rho_{N}=\rho_{0}},
  & \quad K_{sym} = 9 \rho_{N}^{2} \left. \frac{d^{2}E_{sym}(\rho_{N})}{d\rho_{N}^{2}} \right|_{\rho_{N}=\rho_{0}},
    \label{eq:definition-L-Ksym} \\
  J_{0} = 27 \rho_{N}^{3} \left. \frac{d^{3}E_{0}(\rho_{N})}{d\rho_{N}^{3}} \right|_{\rho_{N}=\rho_{0}},
  & \quad J_{sym} = 27 \rho_{N}^{3} \left. \frac{d^{3}E_{sym}(\rho_{N})}{d\rho_{N}^{3}} \right|_{\rho_{N}=\rho_{0}}.
\end{align}

\section{\label{app:qscalardensity} The quark scalar density in matter}

We calculate the isoscalar (isovector) quark scalar density, $G_{\sigma j}$ ($G_{\delta j}$), in the in-medium nucleon, which are defined by
\begin{align}
  \left(\frac{\partial M_{j}^{\ast}}{\partial\bar{\sigma}}\right) &\equiv - g_\sigma G_{\sigma j}(\bar{\sigma}, \bar{\delta}) , 
  \label{eq:qsds} \\
  \left(\frac{\partial M_{j}^{\ast}}{\partial\bar{\delta}}\right) &\equiv - g_\delta G_{\delta j}(\bar{\sigma}, \bar{\delta}) ,    \label{eq:qsdv}
\end{align}
where $j$ indicates proton or neutron.  Because the effective nucleon mass is given by 
\begin{equation}
  M_j^\ast = E_j^{0 \ast} + E^{spin}_j - E^{cm \ast}_j , 
  \label{eq:bmass2}
\end{equation}
where $E_j^{0 \ast}$ and $E^{cm \ast}_j$ are respectively calculated by Eqs.~(\ref{eq:energy1}) and (\ref{eq:Ecm1}) with $\epsilon_i^\ast$ and $m_i^\ast$ ($i=u$ or $d$), 
the derivatives are evaluated as~\cite{saito94ns}  
\begin{align}
  \left(\frac{\partial M_{j}^{\ast}}{\partial\bar{\sigma}}\right) 
  & = - g_\sigma^q \left( \frac{\partial}{\partial m_{u}^{\ast}} + \frac{\partial}{\partial m_{d}^{\ast}} \right) M_j^\ast = - g_\sigma^q \left( \frac{\partial}{\partial m_{u}^{\ast}} + \frac{\partial}{\partial m_{d}^{\ast}} \right)  \left( \sum_{i} \epsilon_i^\ast - E^{cm \ast}_j \right)  \nonumber \\
  & = -(3g^{q}_{\sigma}) \times \frac{1}{3} \left[\sum_{i}S_{i}^\ast - \left(\frac{\partial}{\partial m_{u}^{\ast}}+\frac{\partial}{\partial m_{d}^{\ast}}\right) E^{cm \ast}_j \right] = - g_\sigma G_{\sigma j}(\bar{\sigma}, \bar{\delta})  ,   \label{eq:derivsigma}  \\ 
   \left(\frac{\partial M_{j}^{\ast}}{\partial\bar{\delta}}\right) 
  & = - g_\delta^q \left( \frac{\partial}{\partial m_{u}^{\ast}} - \frac{\partial}{\partial m_{d}^{\ast}} \right) M_j^\ast = - g_\delta^q \left( \frac{\partial}{\partial m_{u}^{\ast}} - \frac{\partial}{\partial m_{d}^{\ast}} \right)  \left( \sum_{i} \epsilon_i^\ast - E^{cm \ast}_j \right)  \nonumber \\
  & = -g^{q}_{\sigma}\left[\sum_{i} \tau_3^i S_{i}^\ast - \left(\frac{\partial}{\partial m_{u}^{\ast}} - \frac{\partial}{\partial m_{d}^{\ast}}\right) E^{cm \ast}_j \right] = - g_\delta G_{\delta j}(\bar{\sigma}, \bar{\delta}) ,   \label{eq:derivdelta} 
\end{align}
where $g_\sigma = 3g_\sigma^q$, $g_\delta = g_\delta^q$, $\tau_3^{u(d)} =1 \, (-1)$ and $m_{i}^{\ast} = m_i - g^{q}_{\sigma}\bar{\sigma} \mp g^{q}_{\delta} \bar{\delta}$ for $\binom{u}{d}$.  Here, using the Hellmann-Feynman theorem, the quark scalar density (without cm correction) can be calculated as 
\begin{equation}
S_i^\ast \equiv \frac{\partial \epsilon_i^\ast}{\partial m_{i}^{\ast}} = \int d{\vec r} \, {\bar \psi}_i({\vec r}\,) \psi_i({\vec r}\,) = \frac{\epsilon_{i}^{\ast}+3m_{i}^{\ast}}{3\epsilon_{i}^{\ast}+m_{i}^{\ast}}  .   \label{eq:qscalar} 
\end{equation}
Next, from Eq.~(\ref{eq:Ecm1}) we can find 
\begin{align}
  \frac{\partial E_{j}^{cm \ast}}{\partial m_{i}^{\ast}} 
  & = \frac{\partial}{\partial m_{i}^{\ast}} \left[ \frac{11}{3}\sum_{k}A_{k}^\ast
  +\frac{1}{9}\sum_{k}\sum_{q} \left( \delta_{kq} - 1 \right) C_{k}^\ast D_{q}^\ast \right]  \nonumber \\
 & = n_{i}\left[\frac{11}{3}A^{\ast \prime}_{i}+\frac{1}{9}C^{\ast \prime}_{i}\left(D_{i}^\ast -\sum_{k}D_{k}^\ast \right)+\frac{1}{9}\left(C_{i}^\ast -\sum_{k}C_{k}^\ast \right)D^{\ast \prime}_{i}\right],   \label{eq:dercm}
\end{align}
with $n_{i}$ being the number of $i$ quark in $j$.  Using Eq.~(\ref{eq:acd1}), we then obtain   
\begin{align}
  A^{\ast \prime}_{i}
  & =\frac{d A_{i}^\ast}{d m_{i}^{\ast}}=\frac{(3S_{i}^\ast -1)(\epsilon_{i}^{\ast2}+m_{i}^{\ast2})+2(S_{i}^\ast -3)\epsilon_{i}^{\ast}m_{i}^{\ast}}{3(3\epsilon_{i}^{\ast}+m_{i}^{\ast})^{2}},  \label{eq:der1} \\
  C^{\ast \prime}_{i}
  & =\frac{d C_{i}^\ast}{d m_{i}^{\ast}}=2(S_{i}^\ast \epsilon_{i}^{\ast}-m_{i}^{\ast})A_{i}^\ast +(\epsilon_{i}^{\ast2}-m_{i}^{\ast2})A^{\ast \prime}_{i},  \label{eq:der2} \\
  D^{\ast \prime}_{i}
  & =\frac{d D_{i}^\ast}{d m_{i}^{\ast}}=\frac{11S_{i}^\ast +1}{(3\epsilon_{i}^{\ast}+m_{i}^{\ast})(\epsilon_{i}^{\ast2}-m_{i}^{\ast2})} \nonumber \\
  & -\frac{11\epsilon_{i}^{\ast}+m_{i}^{\ast}}{(3\epsilon_{i}^{\ast}+m_{i}^{\ast})^{2}(\epsilon_{i}^{\ast2}-m_{i}^{\ast2})^{2}}\left[(9S_{i}^\ast +1)\epsilon_{i}^{\ast2}+2(S_{i}^\ast -3)\epsilon_{i}^{\ast}m_{i}^{\ast}-3(S_{i}^\ast +1)m_{i}^{\ast2}\right].  \label{eq:der3}
\end{align}

Finally, combining Eqs.~(\ref{eq:derivsigma}), (\ref{eq:derivdelta}), (\ref{eq:dercm}) and (\ref{eq:der1}) - (\ref{eq:der3}), we can find the analytical expressions of the quark scalar densities, $G_{\sigma j}$ and $G_{\delta j}$.

\bibliographystyle{apsrev4-2.bst}
\bibliography{DW13233-revised.bib}

\end{document}